# Estimating the Impact of Case Management in MDLs: Lone Pine Orders and Bellwether Trials

Eric Helland[*]
Claremont-McKenna College and RAND

Minjae Yun
Claremont-McKenna College

March 2025

Abstract

Case management by judges is increasingly determining the outcome of litigation, particularly in the multidistrict litigation (MDL) process. One concern is that the MDL process pressures defendants to settle, regardless of the merits, and provides insufficient information on the value of individual cases within the MDL. Critics of the MDL system have suggested two management orders as solutions to these problems. The first is Lone Pine orders, which require plaintiffs in an MDL to produce evidence of injury and causation. The second is bellwether trials, in which the court selects certain cases for trial to provide information on the value of claims and encourage settlement. We examine the impact of Lone Pine orders and bellwether trial processes on the outcomes of cases in multidistrict litigation (MDLs). Using data on MDLs from 1992 to 2017, we find that Lone Pine orders are associated with an increase in the number of cases resolved in the MDL process.

---

[*]The author wishes to thank participants at the RAND Institute for Civil Justice Multidistrict Litigation Conference, CJRI Symposium What's Happening in Federal Court? at Berkeley Law, the CMC Summer Research Conference, the University of Chicago's CELS, the Feinberg Center for Catastrophic Risk Management and Compensation, and comments from Paul Rheingold, James Anderson, Nora Freeman Engstrom, Judge Carolyn Kuhl, Robert S. Peck, Margaret Williams, Emory Lee, Jonathan Nash, and Charles Lifland. The research assistance of Tobin Hansen, Melanie Wolfe, James Dial, Sophia Helland, and Vinodh Srikanth was invaluable. The contents of this paper are solely the responsibility of the author and do not reflect the opinions of the RAND Corporation. Helland can be reached at eric.helland@claremontmckenna.edu.

# 1. Introduction

In the past 20 years, Multidistrict Litigation (MDL) has come to dominate the federal courts' civil docket. The MDL process, which is designed to handle complex cases with similar causes of action, has evolved from a relatively minor portion of the federal caseload in the 1980s to account for almost 40% of the federal civil docket (Resnik, 2016).[1] There are numerous competing explanations for the rise of MDLs, including both legislation and court rulings, which make it more difficult to utilize class actions.[2] However, there is a consensus that the rise reflects a shift away from litigation and toward administrative procedures, at least in the resolution of mass tort claims (See Burch, 2019; Nagareda, 2008). Put differently, the goal of the MDL, and increasingly, all civil litigation in the federal courts, is to settle individual claims before trial. In the case of MDLs, the objective is broader than simply avoiding trials in individual cases. The goal is to achieve comprehensive settlements that resolve entire lines of litigation. As Nagareda (2008) argues, the MDL process is more administrative than traditional tort litigation. It aims to reduce the need for individual adjudication by delivering compensation at a lower administrative cost. To support this administrative resolution, courts have increasingly relied on a range of case management techniques that, while not unique to MDLs, are far more commonly employed in MDLs than in other types of litigation.[3]

---

[1] Multidistrict Litigation is a process specifically designed to handle large and complex cases involving similar injuries or causes of action. The aim is to streamline the litigation by eliminating duplication in pretrial proceedings. For example, in complex litigation involving thousands of cases, witnesses would have to be deposed hundreds of times, creating enormous costs to the courts and the parties. MDLs are created by the U.S. Judicial Panel on Multidistrict Litigation (JPML).

[2] See Willging and Lee (2009). MDLs are not necessarily class actions. A class action is a single case with multiple plaintiffs. In an MDL, the individual cases remain separate claims but are grouped together to resolve pretrial proceedings. At the end of these proceedings, the MDL judge will, theoretically, remand the cases to their original court for individual trials. In Burch's sample of 73 MDLs, 27.3% have at least partial class action settlements.

[3] It is worth noting that the MDL process was not necessarily intended to promote settlement within the MDL (see Bradt, 2018). The MDL statutes direct transferee judges to remand cases for trial upon the conclusion of pretrial proceedings. The vast majority of MDL cases are resolved before the MDL judge remands them (see Burch, 2019).

This paper focuses on two techniques: Lone Pine orders and bellwether trial processes. Lone Pine orders, which were first used in *Lore v. Lone Pine Corporation*, differ in their specific requirements but typically require plaintiffs to provide documentation of exposure and evidence of both general and specific causation by a certain date or risk having their cases dismissed. [4] Proponents of Lone Pine orders argue that by establishing an early threshold to address holes in the plaintiffs' evidence or the presence of meritless claims in the litigation pool, they save both court and defendant resources (see Engstrom, 2020, for a discussion).[5] As we discuss further below, there is an alternative interpretation of Lone Pine orders. They are often part of an effort to resolve an MDL by incentivizing plaintiffs to accept a settlement agreement. According to this view, judges in MDLs perceive their role as facilitating settlement in a timely manner (Burch, 2019; Engstrom, 2019). Lone Pine orders, in Burch and Engstrom's estimation, are part of a broader shift to managerial judging.[6] Of course, these two hypotheses on the use of Lone Pine orders are not mutually exclusive. Some judges could use pre-discovery Lone Pine orders to weed out "meritless" cases, while others could use them to induce settlement.[7]

Bellwether trial processes are orders by the MDL judge to conduct one or more trials. The purpose of the process is to provide a representative sample of all claims in the MDL, which remanded cases may not fully represent. According to the typical justification, these cases will provide information on the value of the pool of cases in the MDL, thereby facilitating settlement.

---

[4] *Lore v. Lone Pine Corp*., No. L-33606-85, 1986 WL 637507 (N.J. Super. Ct. Law Div. Nov. 18, 1986)
[5] There have been several legislative efforts to mandate Lone Pine orders in MDLs, rather than leaving their use up to the judge's discretion. See, for example, The Fairness in Class Action Litigation Act of 2017 and Innocent Party Protection Act. Then Judiciary Committee Chair Bob Goodlatte argued that the bill would adjust the balance between abusive plaintiffs and innocent defendants.
[6] The Lone Pine orders in this dataset are comprehensive for MDLs. Lone Pine orders are used extensively outside the MDL process, and in these instances, they may have significantly different impacts than they do within MDLs.
[7] Engstrom and Espeland (2020) find that 37% of the Lone Pine orders their sample are issued prior to discovery while 25% are "twilight" or "sunset" orders issued after discovery presumably when settlement negotiations are well underway. Engstrom and Espeland have generously made their data public, and it is included in this study.

Because MDL judges are not supposed to conduct trials in the MDL, only resolve pretrial motions in the MDL, MDL judges can only serve as trial judges in cases filed in their specific districts.

In this paper, we examine the impact of Lone Pine Orders and bellwether trial processes on the resolution of cases within multidistrict litigation (MDL). We utilize two datasets: one from the U.S. Judicial Panel on Multidistrict Litigation (JPML), which tracks the aggregate number of cases in the MDL, and another from the Federal Judicial Center (FJC), which examines individual cases. We supplement the FJC data with information from individual case dockets on PACER, which contain indicators of the ultimate resolution of the cases, and the SCALES dataset, which utilizes natural language processing to classify docket entries and case outcomes. In the JPML data, we can examine how rapidly these two case management techniques resolve cases in the MDL. In the case of the aggregate JPML data, we find that a Lone Pine order increases the proportion of cases in the MDL resolved by either settlement, dismissal, or a unilateral drop by the plaintiff by 30%. The impact is almost three times as large as the impact on resolution resulting from plaintiff fact sheets, which involve the disclosure of information without the burden of proving causation. When we control for the potential endogeneity of Lone Pine orders, the impact rises to almost 60%. The evidence on the bellwether trial process is similar, with the bellwether process causing a 13%-25% increase in resolution.

The individual case information in the FJC data provides insight into whether the increase in resolved cases is due to settlements, drops, or dismissals. We consistently find that settlements increase because of Lone Pine orders. We also find that the probability of a unilateral drop by plaintiffs increases after Lone Pine orders are issued. By contrast, the likelihood of dismissal increases after Lone Pine orders only when we recode all cases captured by the SCALES data using the SCALES classification. One issue with interpreting this finding is that while courts appear to be consistent in recording settlements in the data, some courts list cases as dropped even when the case has been settled or dismissed, even when the case is unilaterally dropped, a fact illustrated by our comparisons of the PACER dockets and the SCALES data.

For this reason, a dismissal in a Lone Pine order is not as reliable a measure of the case's merits as it might be in other contexts. It may capture the fact that the plaintiff could not produce the documents rather than indicating that the judge evaluated the merits of the plaintiff's filings in response to the order and found the case to lack merit. The findings are consistent with the theory that Lone Pine orders, and to some extent bellwether trial processes, are part of the MDL judges' toolkit for resolving MDLs, as they seem to push plaintiffs into a settlement. It is more challenging to determine from the available data whether Lone Pine orders result in the dismissal of "meritless" cases. Approximately one-third of the Lone Pine orders in MDLs occur relatively early in the MDL process. However, most of the impact on resolution occurs late in the MDL.

Section 2 of the paper describes the origins of Lone Pine orders and bellwether trial processes. Section 3 describes the datasets used in the analysis. Section 4 presents the methodology, results, and a series of robustness checks. Section 5 concludes the paper.

**2. Calvinball: Background on Lone Pine Orders and Bellwether Trial**

Congress established the Judicial Panel on Multidistrict Litigation (JPML) in 1968 (28 U.S. C 1407). The panel considers cases involving at least one common question of fact pending in different federal districts and determines whether to transfer the cases to a single district court for coordinated pretrial proceedings. If the committee decides that such coordination is warranted, it creates an MDL.[8] As shown in Figure 1, the number of cases in MDLs remains relatively low until the early 2000s. However, since 1996, Congress and the Supreme Court have significantly restricted the use of class actions, and the number of claims handled by the multidistrict litigation (MDL) process has risen dramatically (see Resnick, 2016).

Unlike class actions, which consolidate common cases into a single case, the expressed purpose of an MDL formation, or centralization, is to avoid duplication in discovery, prevent inconsistent rulings on expert evidence, and conserve resources, including the time of judges, lawyers, and parties. Section 1407(b) grants the transferee judge all the transfer court's powers in pretrial proceedings, including pretrial conferences, discovery, motions to dismiss, and summary judgment.[9] A transferee judge can only oversee trials in cases originating in the MDL judge's district or when parties waive objections to venue changes.[10] For this reason, Lone Pine orders and bellwether trial processes operate under somewhat different constraints.

Lone Pine orders and bellwether trial processes arise because MDLs have an ad hoc quality to procedure rulings. Engstrom (2019) quotes a Center for Judicial Studies report that concluded judges had developed very different approaches to dealing with the thousands of cases they faced, a process she refers to as MDL common law. As Engstrom (2019) explains, when one judge innovates in a particularly challenging case, other judges often find the innovation useful in more prosaic settings.

---

[8] In addition, the JPML also selects the district in which the MDL is held and the judge within that district who will oversee the MDL.
[9] After a determination that a case has the same cause of action as the other consolidated cases in the MDL, the case is removed from the court in which the plaintiff filed and transferred to the MDL court. As such, the original judge is referred to as the transfer judge, and the judge overseeing the MDL is the transferee judge.
[10] In *Lexecon Inc. v. Milberg Weiss Bershad Hynes & Lerach, 523 U.S. 26 (1998),* the United States Supreme Court held that the district court conducting the MDL does not have the authority to assign a transferred case to itself for purposes of the trial unless both parties consent to the transfer.

The second feature in understanding Lone Pine orders and bellwether trial processes is that judges in MDLs, perhaps even more than judges in other cases, view their role as resolving the litigation through settlement. For example, Burch (2019) finds that in her sample of product liability and sales practice MDLs, 52.9% of judges actively encouraged settlement, approved private settlements, and enforced private dispute resolutions that resulted from settlement, thereby incentivizing plaintiffs to participate. Burch found that only 8% of judges in her sample took no steps to promote a settlement deal. While one might argue with the specific incentives she identifies as pushing settlements, it is clear judges in MDLs are quite active in promoting settlements. Burch goes so far as to argue that settlements in MDLs share more in common with arbitration than the typical civil case in federal courts.

Plaintiffs and defendants also face pressure to settle. The defense attorneys face settlement pressure because, as with most mass litigation, their clients would like a global resolution even if that resolution comes at a considerable financial cost. Plaintiff attorneys, particularly those serving on plaintiffs' steering committees, face pressure to persuade their clients to accept a settlement. As Burch (2019) notes, courts often link settlement to fees. Even beyond the case-specific financial incentives, being appointed to a plaintiffs' steering committee by a judge in future cases is contingent upon not creating problems for judges who feel their role is to facilitate settlement (see George and Williams, 2013; and Burch, 2019). Given this pressure for settlement, it should not come as a surprise that both Burch (2019) and Engstrom (2019) have argued that Lone Pine orders and bellwether trials can be best understood as innovations to incentivize aggregate settlement.

*2.1 Lone Pine Orders and the Alternatives*

Lone Pine case management orders are those that require plaintiffs to produce evidence to support their claims of harm. The timing and requirements of Lone Pine orders vary across MDLs. Typically, a Lone Pine order's stated purpose is to "identify and cull potentially meritless claims."[11] As Engstrom and Espeland (2020) point out, they have become increasingly common since their first use in 1986.[12]

Theoretically, there are reasons to be concerned with the merits of filings in large MDLs. Engstrom (2019) notes that the litigation dynamics and incentives in the MDL process differ significantly from those in other federal cases. Lawyers in mass MDLs have incentives to sign up as many clients as possible, as the marginal cost of an additional client is often negligible. Filing fees may be waived for direct filings into the MDL[13], and common discovery, expert reports, and bellwether structures reduce per-case expenses. In contrast, in non-MDLs, each additional client typically requires bespoke discovery, experts, and attorney time. MDLs consolidate these costs, so adding an extra plaintiff incurs a minimal incremental burden. As Rheingold (2019) observes, the cost of acquiring clients has also plummeted with the advent of internet advertising. Consequently, lawyers in MDLs have fewer incentives to screen for case quality, especially in pharmaceutical or environmental MDLs, where defendants often request Lone Pine orders to combat weak claims—though such orders remain rare outside these contexts.

---

[11] Baker v. Chevron USA, Inc., No. 05-227, 2007 WL 315346, at *1 (S.D. Ohio Jan. 30, 2007) cited in Engstrom and Espeland (2020) footnote 55.

[12] Although the first court used Lone Pine in 1986, Judge Fallon's Lone Pine order in the Vioxx MDL helped kick-start the use of Lone Pine orders in mega MDLs (Burch, 2019).

[13] In standard federal civil litigation, a filing fee is required for each new case—currently around $400 in district courts. However, in MDLs, this requirement may be bypassed or modified in certain circumstances. For example, when a plaintiff files a case directly into the MDL using a short-form complaint or through an administrative docket process, courts may treat the filing as part of the centralized proceeding rather than as a new civil action, thus avoiding the need for a separate fee. Additionally, some MDL courts authorize the use of master complaints and direct filing orders that streamline intake and reduce per-plaintiff costs, including filing fees. These practices are not uniform across MDLs and typically depend on the structure authorized by the transferee judge. See Manual for Complex Litigation (Fourth) § 20.131 (2004).

Lone Pine orders typically require plaintiffs to provide evidence of exposure to the product or toxin at issue in the case, along with specific details about the nature of the exposure (e.g., taking a medication in a pharmaceutical case). Second, the plaintiff must demonstrate harm, and finally, the plaintiff must provide evidence of specific causation in the form of a report from an expert witness (See Engstrom and Espeland, 2019; Burch, 2019). One might argue that Lone Pine orders are a straightforward way to eliminate claims with dubious merit from the case pool, as showing exposure, harm, and causation are essential elements of a successful case. However, Engstrom (2019) argues that the provision of medical records itself is not costless. Moreover, Lone Pine orders often require proof of both general and specific causation, thereby placing an evidentiary burden on plaintiffs earlier in the litigation process. A second concern, raised by Engstrom (2020) and Burch (2019), is that Lone Pine orders are not part of federal rules; as such, there is limited guidance on how judges should implement them.

There is both public policy and judicial interest in the use of Lone Pine orders. For example, in 2018, Congress considered H.R. 985, which would have required plaintiffs in MDLs to submit evidence of injury within 45 days of filing or transfer. This proposed use of a Lone Pine order differs sharply from how such orders are typically employed in MDLs. Burch (2019) argues that Lone Pine orders generally occur much later in the MDL process than commonly assumed. Rather than serving as early gatekeeping devices, they often function as "nudges" by MDL judges to encourage reluctant plaintiffs to accept a settlement, usually one negotiated by the plaintiffs' steering committee. In the Fosamax litigation, Burch quotes plaintiff leadership describing Lone Pine orders as a "post-settlement mop-up procedure". At the same time, defendants often frame them as "put up or shut up" orders. Engstrom (2020) makes a similar point, characterizing them as "twilight orders" that force plaintiffs to either accept settlement terms or face an accelerated evidentiary burden and potential dismissal.

Our data are consistent with a distinction between two types of Lone Pine orders used in MDLs. The first category, which we term early-stage Lone Pine orders, appears aimed at screening out weak claims before settlement discussions begin. These orders require plaintiffs to submit evidence of exposure, injury, or causation and are most commonly used in cases where fraudulent or unsupported claims are a concern. In the *Silica* litigation (MDL 1553, 2005), a Lone Pine order was issued after courts discovered widespread fraudulent medical screenings supporting silicosis claims. In *Vioxx* (MDL 1657, 2007), the judge implemented a Lone Pine order before settlement negotiations to ensure that only plaintiffs with documented medical injuries remained in the litigation. Similarly, in the *Deepwater Horizon Oil Spill* (MDL 2179, 2014), given the massive volume of claims, many of which lacked supporting evidence, the court used a Lone Pine order as an initial screening mechanism.

However, the more common type of Lone Pine order in our data aligns with those characterized by Engstrom and Burch. These orders appear to be designed to encourage plaintiffs to settle by requiring them to produce specific expert reports within a short timeframe or risk having their case dismissed. In *Fosamax* (MDL 1789, 2012), as cited by Burch, the judge issued a Lone Pine order after bellwether trials and a global settlement framework, compelling lingering cases to either provide proof or settle. A similar pattern emerged in the *Zimmer NexGen Knee Implant* litigation (MDL 2272, 2016), where the court issued a Lone Pine order in response to repeated dismissals of bellwether cases, leading to mass case dismissals, attorney withdrawals, and inventory settlements. [14] In some cases, such as *Testosterone Replacement Therapy* (MDL 2545, 2018) and *Xarelto* (MDL 2592, 2019), the transferee judges issued Lone Pine orders only after global settlement agreements had been reached. Lone Pine orders are also used as cleanup mechanisms in MDLs without a global settlement. For instance, in *CR Bard Pelvic Repair Systems* (MDL 2187, 2018), the judge issued a Lone Pine order after most inventory settlements had been finalized, requiring remaining plaintiffs to either submit evidence of injury or exit the litigation.[15]

There are less burdensome alternatives to Lone Pine orders for assessing whether a plaintiff's case meets a minimal evidentiary threshold. Specifically, Williams et al. (2019) and Engstrom (2020) identify Plaintiff Fact Sheets (PFS) as a common alternative. PFS are standardized questionnaires that function like interrogatories and requests for production of records, requiring plaintiffs to provide basic case information such as product use history, medical records, and alleged injuries. The critical distinction between PFS and Lone Pine orders is that PFS does not require proof of causation, and judges typically do not dismiss cases solely for missing the PFS deadline, though repeated noncompliance can result in sanctions.

Like Williams et al. (2019), we also identify Plaintiff Profile Forms (PPF) in several MDLs. Williams et al. distinguish PPFs from PFS, noting that PPFs are even less detailed and serve primarily as preliminary screening tools. For completeness, we also consider Defendant

---

[14] As we discuss further below, there is a relationship between bellwether processes and Lone Pine orders. In *In re: Zimmer,* the impetus for the judge's Lone Pine order appears to have been the plaintiffs' voluntary dismissal of 15 of 16 bellwether trial cases—presumably to avoid adverse verdicts—and their continued pursuit of claims that the defendant argued had little evidence of causation. The court issued a Lone Pine order in June 2016, requiring plaintiffs to submit expert affidavits supporting specific product defect and causation claims. Many plaintiffs failed to comply; attorneys withdrew from cases, and over 1,400 of the more than 1,700 cases were ultimately dismissed or withdrawn. The remaining cases were resolved through confidential inventory settlements, which were announced in early 2018. See Case Management Order No. 11, No. 11-cv-5468 (N.D. Ill. June 10, 2016)

[15] The nudge becomes more forceful to the extent that the settlement agreement encourages lawyers to recommend settlement to all their clients and subsequently withdraw from representing any client who declines to participate in the settlement program (Burch, 2019).

Fact Sheets (DFS), which courts order defendants to complete by providing relevant information about plaintiffs already in their possession, such as sales data, customer complaints, or internal testing records. These case management tools are not mutually exclusive. Many MDLs employ various combinations of PFS, PPF, DFS, and Lone Pine orders at different stages of litigation, depending on the complexity of the claims and the judge's approach to case management.[16]

While both Engstrom's (2019) and Burch's (2019) views of Lone Pine orders, i.e., screening devices for which there might be less burdensome alternatives (Engstrom) or nudges by which judges can incentivize reluctant plaintiffs to participate in a settlement (Engstrom & Espeland (2019) and Burch (2019)), are not necessarily mutually exclusive, they do represent an unresolved empirical question. Across Lone Pine orders, do these orders lead to an increase in the fraction of cases resolved in the MDL, and do Lone Pine orders increase the likelihood of settlement, dismissal, or drops, or all three?

Before turning to bellwether trials, a data issue arises in examining the resolution of cases using both the JPML data and the FJC data discussed below. The JPML data provides an annual count of the number of cases in the MDL and the aggregate number of cases resolved (presumably via settlement, dismissal, or unilateral drop by the plaintiff, i.e., abandoned by the plaintiff without being dismissed by the judge for a cause such as failure to comply with a Lone Pine order). As we discuss further below, the FJC data can be used to distinguish between settlement, dismissal, and drops. However, the meaning of 'dismissal' and 'drop' in the context of Lone Pine orders is unclear. The FJC data do not distinguish between voluntary dismissal or drop and judge dismissal for failure to comply, meaning that a drop and dismissal may have different meanings across different courts. This limitation does not affect the ability to test the hypothesis that Lone Pine orders serve as nudges for settlement versus screening for weaker cases. It prevents us from examining whether Lone Pine orders are primarily impacting 'meritless' cases, as we cannot distinguish between cases in which the judge has ruled on the merits and those in which the plaintiff simply could not comply with the Lone Pine order for reasons other than the case being meritless. To examine this data limitation more closely, we have scraped the individual case dockets from PACER to identify indicators that would allow us to differentiate between different case outcomes. Even this data has inherent limitations, as it is unclear whether such data exists in all cases, since law firms often keep track of Lone Pine compliance and simply inform courts that cases have been dropped.

---

[16] There is a relationship between plaintiff fact sheets and bellwether processes. PFS are common in cases where judges have utilized bellwether trials, perhaps because they have adopted early electronic plaintiff fact sheets, which facilitate the creation of case selection pools.

*2.2 Bellwether Trials*

While there is some debate in the academic literature about the purpose of Lone Pine orders, i.e., whether they are primarily a screening device to weed out meritless cases or a judicial tool for incentivizing settlement, the rationale for bellwether trials, according to the far more extensive academic literature on bellwether trials, is more straightforward. The stated purpose is to have a small group of claims that serve as a litmus test for potential future trials, thereby informing settlement negotiations.[17] Of course, the value of bellwether trials in settlement negotiations hinges on whether the process produces reliable information about the case pool. There are numerous methods that MDL judges have employed in selecting cases for bellwether trials (see below). However, whether bellwether trials fulfill this promise remains a matter of debate. Fallon et al. (2008) describe bellwether trials as opportunities to develop trial packages, improve litigation efficiency, and precipitate settlements through representative verdicts. However, Lahav (2008) cautions that they risk undermining the decentralization of jury trials and may introduce systemic bias when only a small set of juries shape outcomes for thousands of claims. There is little systematic evidence on their impact on settlement probability.

---

[17] In its guide to bellwether trials, the Federal Judicial Center defines a bellwether trial as "individual trials conducted by MDL transferee judges to produce reliable information about other cases centralized in that MDL proceeding" (Whitney 2019, 3).

There are several reasons to fear that, even if the process generates a representative sample of cases from the pool of cases in the MDL, the eventual trials will not be random; hence, it is far from clear that they produce unbiased information about claim value in the MDL. Both sides have an incentive to shape the pool of cases eventually litigated. Defendants may aggressively settle unfavorable cases, and plaintiffs often drop weak cases selected for a bellwether trial (see Brown et al., 2014).[18] In light of the selection effects, even with a process that would produce a random selection of cases, the actual cases litigated are unlikely to be a random selection of cases. Both sides can make cases unfavorable to them go away.

The second issue is that not all cases in the MDL may be eligible for inclusion. In *Lexecon Inc. v. Milberg Weiss*, the Supreme Court held that a transfer judge could not "self-transfer" an MDL action to their district to conduct a bellwether trial unless the litigants waived their right to the original venue. It is likely that a large number of cases, which may be very distinct in terms of their facts, would not be eligible for the bellwether process unless, under the Lexecon decision, the parties involved consent to the transfer.

---

[18] Whitney cites the Duke Mass-Tort Practices guide in arguing that this selection might be beneficial,

> *Many bellwether cases resolve along the way, whether because of errors in the plaintiff fact sheet, special factors that strengthen or weaken the case during discovery that were not anticipated at the outset, or because of the court's early rulings. These cases should not be regarded as failures. Instead, they are important data points, helping the lawyers better understand the ground reality of the cases—which may vary considerably from the hypothetical plaintiff that has been the idealized subject of early negotiations. Indeed, the reasons these cases drop out—gamesmanship, good advocacy, plaintiffs disappearing, the outcome of preliminary motions—all provide insights into how the broader pool of cases may fare."* Whitney (2019) page 31 footnote 52

A final issue for estimating the impact of the bellwether trial process on case resolution is that judges have used a wide range of selection methods. The FJC's pocket guide recommends choosing bellwether trials through random selection.[19] Far more common are processes in which both sides each select their best cases from a pool of potential plaintiffs.[20] [21] The list approach is as far from random as possible, given that both parties have an incentive to present outlier cases that favor their position.[22] Many bellwether processes have far more complex processes for case selection.[23] Anecdotally, the success of bellwether trials in producing settlements depends critically on the judge running the process. The process itself may serve less to inform the settlement negotiations about the value of cases within the pool of claims in the MDL, but rather serve as another way to facilitate settlement. Put differently, creating a bellwether process is another way for the judge to signal she is serious about resolving the MDL.[24]

Put differently, the argument is that while bellwether trials were designed to resolve uncertainty through representative adjudication, in practice, they more often function as institutional tools for facilitating and legitimizing settlements. The success of this process depends not just on case selection but on the judge's role in managing information flows, pretrial rulings, and party coordination. The bellwether process is less about adjudicating individual cases and more about generating information for private settlement negotiations.

---

[19] For example, Lahav (2008, 2018) argues that bellwether trials should be selected through random sampling to ensure representativeness and reduce strategic manipulation. Random selection, she contends, would enhance the legitimacy of bellwether outcomes and provide more reliable information to guide global settlement negotiations. Fallon et al. (2008) caution that party-selected bellwether cases often reflect each side's strategic preferences rather than the broader case population, thereby undermining their predictive value.

[20] For example, in the *Xarelto* MDL, Judge Fallon allowed the defense to select 10 cases, the plaintiffs to select 10 cases, and Judge Fallon then selected the remaining 20 cases. Four bellwether cases were then randomly selected.

[21] In the *FEMA trailer* MDL, which contained 60,000 cases, the court narrowed the cases available for bellwether before allowing plaintiffs to choose the cases. The court selected four bellwethers with trial dates set for February 2009 and January 2010. First, the plaintiffs' lawyers nominated individual cases. Next, the court selected four test cases —one for each of the manufacturer defendants.

[22] In our sample of MDLs, over half involve selection by the plaintiff and defense, while only 27% involve some element of random selection.

[23] For example, in the *General Motors Ignition Switch* MDL, the plaintiffs, defendants, and the court identified bellwether cases for the initial discovery pool. The court then chose a random subset of cases from the list for additional pretrial discovery. In the Medtronic Implantable Defibrillators MDL, the court required each party to identify no more than six categories for bellwether trials, assign each plaintiff to one of these categories, and then randomly select bellwethers from the established categories. This process was followed by peremptory challenges from each side to reduce the number of bellwethers to three.

[24] Zimmerman (2017) argues that in practice, bellwether trials rarely yield generalizable information and often morph into what he calls "bellwether settlements"—mediated agreements based on a representative sample of settlement values rather than jury verdicts. These settlements avoid the high stakes and unpredictability of jury trials, offering instead a structured, iterative process that builds trust among lawyers and facilitates global resolution.

**3. MDL Data Sets**

*3.1. Data Description: JPML and FJC data*

To measure the impact of case management innovations on case resolution, we utilize two data sets. The first is from JPML. Each year, the JPML publishes the Statistical Analysis of Multidistrict Litigation, which has case counts by year by MDL.[25] The JPML categorizes cases as closed, remanded, or pending, enabling us to calculate the percentage of cases resolved each year (closed + remanded / total). The data do not indicate how the cases were resolved, i.e., whether they were settled, dismissed, or unilaterally dropped. This data's main advantage is that it contains a complete count of all cases in the MDL. The JPML data covers the period from 1992 to 2017 and includes 1,335 total MDLs, of which the JPML classified 363 as involving personal injury claims, primarily pharmaceutical cases.

The second data source is the Federal Judicial Center's Integrated Database. The FJC produces the data in cooperation with the Administrative Office of the U.S. Courts (AO).[26] The data includes information on civil case filings and terminations, MDL identifiers, and case outcomes, categorized under various AO disposition codes. Crucially, these codes are used to approximate case resolution types, such as settlement, dismissal, or dropped cases.[27]

---

[25] See, for example, https://www.jpml.uscourts.gov/sites/jpml/files/JPML_Statistical_Analysis_of_Multidistrict_Litigation-FY-2017.pdf.
[26] See https://www.fjc.gov/research/idb.
[27] The codebook identifies the variable as "the manner in which the case was disposed of." The options available are Cases transferred or remanded:
    0 – transfer to another district
    1 – remanded to state court

However, as Eisenberg & Lanvers (2009), Claremont (2009), and Hadfield (2004) emphasize, defining settlement is not straightforward in the FJC/AO data. Some studies adopt a broad definition of settlement that includes any resolution without adjudication, even those in which the plaintiff recovers nothing, such as voluntary dismissals or dismissals for want of prosecution. Others use a narrower definition, limiting settlements to those with a clear, consensual resolution between the parties, which theoretically results in a payment from the defendant to the plaintiff. In our analysis, we use a narrower approach, categorizing AO Code 13 (dismissed: settled) and Code 5 (judgment on consent) as settlements. Unlike Hadfield's broader approach, we classify Codes 12 and 14 (voluntary and other dismissals) and Code 2 (dismissed for want of prosecution) as dropped cases, and dismissed cases as Codes 3 (dismissed for lack of jurisdiction), 4 (dismissed due to default), and 6 (judgment on motion before trial).

In multidistrict litigation, the meaning of "settled" can vary significantly depending on whether the resolution occurs through an inventory settlement or a global settlement. In an inventory settlement, the plaintiff's attorney negotiates a bundled resolution with the defendant that covers some or all of the cases in their firm's portfolio, often based on informal categorizations of injury severity or product exposure. When a case is marked as "settled" in the FJC/AO data and, as shown below, it typically has the same entry in the PACER docket, it usually means that the client has agreed to the terms, the necessary paperwork has been executed, and payment is either complete or imminent.[28] By contrast, in a global settlement—such as in the NFL Concussion MDL or the Stryker Rejuvenate and ABG II hip implant litigation—the docket entry "settled" indicates that the case has entered a claims administration phase. At that point, settlement approval and payment depend on the claims administrator's evaluation of whether the claim meets the terms of the settlement program, including proof of product use, injury, causation, and eligibility under a predetermined compensation grid. In such cases, many claims may ultimately be dismissed or withdrawn if they fail to meet the settlement criteria.[29]

---

        10 – multidistrict litigation transfer
        11 – remanded to U.S. Agency
Dismissals:
        2 – want of prosecution
        3 – lack of jurisdiction
        12 – voluntarily
        13 – settled
        14 – other
Judgment on:
        4 – default
        5 – consent
        6 – motion before trial
        7 – jury verdict
        8 – directed verdict
        9 - court trial
        15 – award of arbitrator
        16 – stayed pending bankruptcy
        17 – other
        18 – statistical closing
        19 – appeal affirmed (magistrate judge)
        20 – appeal denied (magistrate judge)

The key problem for this study is that courts differ on what they record as dismissed or dropped. For example, a case that does not comply with a Lone Pine order would typically be dismissed by the courts for want of prosecution. Nevertheless, some courts list these cases as dropped because the plaintiff failed to provide the necessary documentation, i.e., due to a lack of prosecution on their part. For this reason, we treat drops and dismissals independently.[30]

---

[28] The pelvic mesh litigation against C.R. Bard, consolidated in *In re C.R. Bard, Inc., Pelvic Repair System Products Liability Litigation*, MDL No. 2187 (S.D. W. Va.), provides an instructive example of resolution through inventory settlements. Rather than entering into a single global agreement, Bard negotiated confidential settlements with individual plaintiffs' firms, often resolving large portfolios of cases based on the type of injury, treatment history, and product usage. These settlements were administered privately, with the court overseeing the broader MDL structure but not directly managing claim validation or payout. Once plaintiffs opted in and met documentation requirements, their individual dockets were typically marked as "settled," reflecting final resolution and movement toward payment.

[29] In the NFL case, the court approved a global settlement that established a claims administration system overseen by a claims administrator, requiring players to submit documentation of qualifying diagnoses tied to cognitive impairment or neurological disease. Claims are evaluated for eligibility, causation, and fraud risk, and payment amounts are determined by a publicly available matrix based on diagnosis, age, and years played. See *In re National Football League Players' Concussion Injury Litigation*, MDL No. 2323 (E.D. Pa.), Final Approval Order, ECF No. 6534 (Apr. 22, 2015). Similarly, the Stryker Rejuvenate and ABG II hip implant MDL (MDL No. 2441) was resolved through a global settlement, which allowed claimants who had received surgery to participate in a program featuring a published base award and enhancement framework. The settlement included an eligibility verification phase, product identification, and medical criteria, all of which were met prior to payment being made. See *In re Stryker Rejuvenate and ABG II Hip Implant Products Liability Litigation,* MDL No. 2441 (D. Minn.), Master Settlement Agreement (Nov. 3, 2014). In both cases, once a case is marked as "settled" on the docket, it typically signals a transfer to the claim's administrator, not immediate payment, subject to further scrutiny and validation.

The MDL number was frequently missing in the FJC data prior to 2005; therefore, we limited the FJC sample to the period from 2005 to 2019. We have merged the FJC data with a sample of transfer orders from the personal injury MDLs, and the MDL number is present in over 90% of the cases. If an MDL transfer order identifies the case, the FJC data almost always has an MDL number. However, the FJC data appears to undercount the number of resolved cases relative to the JPML data, possibly due to direct filings into the MDL.[31] The typical MDL in the FJC data "misses" the number of resolved cases. The concern is that this failure to capture cases might be systematic. For this reason, we estimate the model with both the FJC/AO data and the more limited but comprehensive JPML data. If, as turns out to be the case, both data sources yield a similar finding —i.e., an increase in resolutions due to the initiation of a Lone Pine order or bellwether process —we have more confidence that the missing cases are not unduly influencing the results.[32]

*3.1.1 Supplementing AO/FJC Disposition Codes with PACER Docket Data*

To supplement the somewhat coarse case termination codes provided in the FJC/AO data, we incorporated additional information scraped from PACER dockets through the CourtListener

---

[30] More broadly, it is unclear whether dropped cases have had a decision on the merits, as would be the case in a summary judgment motion, or if the plaintiffs simply abandoned their case, which we partially verify with our sample of PACER dockets.

[31] Although this is not completely clear. The FJC/AO data shows a spike in filings at the district office where the MDL is transferred once the judge allows direct filings into the MDL, suggesting that the FJC/AO data captures some direct filings.

[32] Both the FJC and JPML data have one further limitation. Neither dataset contains information on settlement terms. Both Burch (2019) and Rhinegold (2006) have extensive data on the types of settlements, particularly in larger, multi-district litigation (MDL) cases involving personal injury. However, their information is not at the individual level, so we cannot determine the exact nature of the settlements that individuals might enter into due to the initiation of a Lone Pine order or bellwether process using either FJC or JPML data.

RECAP Archive, a free and open database maintained by the Free Law Project. PACER provides detailed docket entries and case filings. While the AO/FJC database assigns standardized disposition codes to each terminated case, the corresponding PACER docket entries often contain more specific language about the nature of a case's resolution.[33]

From the PACER/RECAP scraped data, we label a case as "settled" if it was classified as settled in the FJC data or if the PACER docket includes phrases such as "stipulated dismissal" or "case settled." As shown in Figure 2, the majority of cases are listed as settled in the FJC data. When we modify this classification based on the PACER dockets, we find a small number of cases moving from the dismissed category to the settled category. However, primarily, we find that cases classified as dropped have a settlement indicator on the docket.

Interestingly, for cases classified as dismissed in the FJC/AO data, we find a subset in which the docket explicitly lists the case as dismissed for want of prosecution or due to judicial orders for noncompliance with case management obligations. For example, we find entries noting that cases are dismissed for failure to comply with a plaintiff fact sheet, Lone Pine orders, or a broader case management order. We also classify cases resolved through summary judgment or judgment on motion before trial as dismissed when there is no indication of a settlement on the docket, as these types of dispositions generally reflect a one-sided resolution on the merits or procedural grounds without the plaintiff's consent. The big picture, however, is that explicit dismissals are rare in the FJC/AO data, and even so, we are likely classifying several drops as dismissals.

---

[33] Nearly all individual cases transferred into an MDL have a corresponding PACER docket, though the level of detail varies. Once transferred, the extent to which filings continue in the individual case docket, as opposed to being consolidated into the MDL's master docket, depends on the practices of the transferee judge. Some courts centralize all filings within the master docket, while others allow or require continued filings in individual dockets. As a result, many MDL-related individual dockets contain only a handful of entries, often limited to the original complaint, the transfer order, and a final resolution entry such as a dismissal or settlement. Nonetheless, these abbreviated dockets typically include sufficient information to determine the outcome of the case, particularly when the case is dismissed, voluntarily dropped, or resolved through settlement. These records are accessible through PACER and often archived in the CourtListener RECAP database, a publicly available repository maintained by the Free Law Project. Although RECAP coverage is not uniformly comprehensive, it is extensive for high-volume MDLs and cases resolved after 2012, when RECAP adoption became widespread. One problem with using CourtListener is that RECAP relies on individual users who access PACER through a browser extension that captures and uploads dockets and filings to the public archive. As a result, CourtListener's coverage is limited to documents that have been previously downloaded by users with the RECAP extension and is not comprehensive. The benefit, from our perspective, is that PACER documents cost ten cents per page, and the typical individual case in an MDL typically has one or two pages of docket information. This suggests that the cost of capturing all 284,206 cases in our post-2012 sample could be over $28,000.

Finally, we classified cases as dropped if the PACER docket indicates voluntary dismissal or entries signaling that the plaintiff's attorney withdrew without replacement, which almost always results in a dismissal in the FJC classifications. In line with Eisenberg and Lanvers's caution about misclassifying silent settlements as drops, we only classify cases as dropped when there is no mention of settlement or stipulation in the docket. By contrast, if the docket indicates both a withdrawal and a settlement, we classify the case as settled. The PACER data is generally available after 2012. Of the 284,206 cases in the PACER/AO data file from 2012 onward, we can match 183,414 (64%), which is sufficient to estimate the model using only the matched data.

*3.1.2 Supplementing AO/FJC Disposition Codes with SCALES Data*

Our second robustness check on our classification system involves merging our data with the SCALES data. The SCALES project systematically extracts and analyzes data from federal court dockets, utilizing natural language processing to classify litigation events (Alexander et al., 2024). Rather than relying solely on AO/FJC disposition codes, SCALES also utilizes docket text, including minute entries, orders, and filings, scraped from PACER, to identify motions, rulings, settlements, and terminations. Since SCALES specifically classified cases as dropped, dismissed, or settled, we replace the FJC/AO identification with the SCALES identification if the case is matched in our sample. The SCALES data is comprehensive for cases filed in 2016 and 2017, although we can match cases as far back as 2001. If we take the sample from 2005, we match 105,271 of the 380,075 cases in our sample, which is 27%.[34]

When we compare the FJC/AO classifications to the SCALES classification, we again find that the proportion of cases classified as settlements is relatively unchanged. A smaller proportion than our PACER classification is reclassified as drops. However, the majority of the movement between the two classification schemes is again between dropped and dismissed cases, suggesting that dismissals and drops in MDLs are difficult to distinguish.

With both the PACER and SCALES data, we do not have a sufficiently long time series to estimate our model without relying on the FJC/AO data to classify cases not covered by the respective sample periods. For this reason, our robustness check should be thought of as testing whether the results are robust to a reclassification of a portion of the data.

*3.2 Lone Pine, Fact Sheet, and Bellwether Data*

The data on Lone Pine orders, bellwether trial processes, plaintiff fact sheets, plaintiff profile forms, defense fact sheets, and data on summary judgment motions, Daubert motions[35], judicial approval of settlements, and judicial approval of special masters comes from a variety of sources. In this study, we follow Williamson et al. (2019) in defining a plaintiff fact sheet as "standardized questionnaires that serve the same function as interrogatories and requests for production" (Williamson et al., 2019, 2). They differ from Lone Pine orders in that they do not require plaintiffs to provide expert evidence establishing causation. Plaintiff profile forms are less extensive than plaintiff fact sheets, and courts generally use these terms to describe less comprehensive questionnaires than plaintiff fact sheets (Williamson et al., 2019). Finally, defendant fact sheets are "questionnaires ordered by the court to collect information about plaintiffs that is in the defendant's possession, or, in some instances, to collect information about defendants" (Williamson et al. 2019, 3). It is worth noting that in our sample, all of the defendant fact sheets collect information on plaintiffs that is in the defendant's possession, and none appear to collect information on the defendant.

---

[34] For cases filed in 2016 and 2017, the match rate is much higher, at 79,252 out of 81,385, or 97%.
[35] Daubert motions address the admissibility of expert testimony. Given the critical role expert testimony plays in establishing both general and specific causation in MDLs, judicial rulings on Daubert motions often serve as pivotal junctures in these proceedings (see Helland, 2019).

Both Engstrom (2020) and Burch (2019) have collected data on Lone Pine orders and made their datasets publicly available. [36] *The Drug and Device Law* blog, particularly posts by contributor Michelle Yeary, also regularly publishes updates on Lone Pine orders. [37] In addition, we collected docket sheets for the MDLs from PACER (as opposed to individual case dockets used to classify case outcomes) and searched them to identify Lone Pine orders. These search parameters were developed based on the Lone Pine orders identified by the sources above. We followed a similar procedure to identify bellwether trial processes, drawing from Burch's dataset and additional cases identified by Whitney (2019). For each bellwether process, we located and reviewed the relevant case management orders in the MDL dockets on PACER. Burch also provides information on other case management tools, such as fact sheets and profile forms, which we used to build additional search terms for PACER. We supplemented her data with information from Williams et al. (2019) concerning plaintiff fact sheets, plaintiff profile forms, and defense fact sheets. In addition, we searched the PACER MDL dockets for indicators of the other case Burch's case management tools. Finally, we searched publicly available sources for information on whether cases in the MDL were resolved via a global settlement or an inventory settlement. The final dataset includes 31 Lone Pine orders and 38 bellwether trial processes (see Tables 1 and 2).

---

[36] Engstrom's sample is available at https://lonepineorders.law.stanford.edu/, and Burch's data is available at https://mdl.law.uga.edu/.
[37] See, for example, https://www.druganddevicelawblog.com/2012/11/lone-pine-cheat-shee.html.

*3.3 Preliminary Data Analysis*

This paper's central question is the impact of two of the more controversial case management tools in MDLs: Lone Pine orders and bellwether trial processes. In Figure 3, we examine the average fraction of total cases in an MDL that are resolved by year since centralization. Not surprisingly, the proportion of resolved cases increases as the MDL has been active for a longer period. Most MDLs conclude within five to six years, so the number of MDLs with pending cases decreases as we move to the right of the figure. The upper panel breaks the time series into periods with and without a Lone Pine order in effect. Although the error bars indicate a fair amount of noise, for periods more than five years after centralization, MDLs operating under a Lone Pine order appear to have a higher proportion of resolved cases. The lower panel shows a similar pattern. It separates periods in which a bellwether process is active from those in which no such process has been initiated. Notably, in the case of Lone Pine orders, the results suggest that early use of these orders has little impact on resolutions. In contrast, orders issued later in the MDL's life cycle are associated with a greater proportion of resolved cases.

Given the noisiness of the data, these figures should be interpreted as suggestive rather than definitive. Still, they indicate that Lone Pine orders and bellwether processes may be associated with increased resolution rates within MDLs. In the next section, we estimate a more complex model to assess whether the patterns observed in Figure 3 remain robust when control variables and MDL fixed effects are included. We then turn to case-level data from the FJC to examine whether these procedural tools influence the likelihood of settlement, dismissal, or voluntary case withdrawal.

## 4. Estimation Method and Results: JPML data

*4.1 Fractional Response Probit*

In the previous section, we examined the proportion of cases resolved each year in the JPML data depending on whether the MDL had a Lone Pine Order or Bellwether process in effect during the year. In this section, we estimate a fractional response probit model to determine whether the proportion of cases resolved is influenced by either of the two case management orders while holding constant the time-invariant characteristics of the MDL via MDL fixed effects. The estimating equation for the JPML data is

$$y_{it} = \alpha_i + \beta Lone\ Pine_{it} + \gamma Bellwether_{it} + X'_{it}\delta + \gamma_t + \epsilon_{it}$$

where $y_{it}$ is the proportion of resolved cases and $\epsilon_{it}$ is the robust error term clustered on MDL. In the JPML data, we cannot differentiate between the different outcomes as the JPML only records that the cases were resolved without being remanded and/or going to trial.

$Lone\ Pine_{it}$ and $Bellwether_{it}$ are indicators for an ongoing Lone Pine order or bellwether process, respectively. *X* are the policy variables and includes controls for plaintiff fact sheets, plaintiff profiles, and defendant fact sheets being in effect. *X* also includes controls for the number of cases in the MDL, the number of cases that have gone to trial in the bellwether process, and the fraction of those cases in which the plaintiffs have won. The controls also include whether the MDL was certified by the judge as a class action, whether the judge ruled on summary judgment motions, a Daubert motion, whether there was a judicially approved settlement, or whether a judicially approved special master was appointed $\gamma_t$ are year fixed effects, and $\alpha_i$ are MDL fixed effects.

A key issue is that we have an unbalanced panel, where each MDL has its own intercept. $\alpha_i$. Moreover, MDLs vary in their duration, with some resolving in one or two years and the Asbestos MDL (875) lasting multiple decades. Moreover, since $y_{it}$ is the fraction of cases in the MDL disposed, it is bounded by zero and one inclusively, so the model is estimated as a fractional response model, i.e., $y_{it}$ satisfies $0 \leq y_{it} \leq 1$. The data are not censored, but in specific years there are corner solutions such as $y_{it}$=0, no cases in the MDL resolve in year *t*, or $y_{it}$=1, if the case reaches a global settlement. One issue with the estimation is that linear panel data models are not well-suited for measuring the fraction of cases (See Papke and Wooldridge, 1996, 2008). Linear panel data models are not bounded by zero or one; therefore, a linear model is, at best, an approximation and may miss important nonlinearities. In fact, given the above discussion of the uses of Lone Pine orders and bellwether trials, it seems likely that edge solutions in the final year of the MDL are important to estimating their effects.

Papke and Wooldridge (1996) and Wooldridge (2010) provide a methodology for estimating fractional response models using a generalized linear model. In the context of the MDL data, it seems likely that there are unobserved but constant characteristics of the different MDLs, necessitating the inclusion of MDL fixed effects. Papke and Wooldridge (2008) demonstrate how to estimate fractional response models for panel data, and Wooldridge (2011) generalizes the estimation technique to accommodate unbalanced panels.

One further concern is that the specific MDLs and perhaps the timing of Lone Pine orders and bellwether trial processes within the MDL may not be exogenous. Our empirical approach addresses these concerns in several ways. First, rather than comparing average settlement or resolution rates across MDLs, we estimate changes in the proportion of resolved cases within MDLs over time, conditional on the stage of the proceeding. Both the fractional response models and the discrete-time hazard models exploit within-MDL variation, and we include MDL fixed effects to absorb unobserved, time-invariant heterogeneity across proceedings (e.g., litigation type). We also include flexible controls for how long ago the JPML centralized the case, i.e., how long the MDL has been going on. These controls help mitigate concerns that observed effects reflect the natural progression of MDLs toward resolution. The remaining concern is that judges who issue Lone Pine orders or implement bellwether processes may also engage in other unobserved forms of active case management when litigation reaches its final stage, which could confound estimates (see Helland and Yun, 2023).

There is no obvious solution to this problem. If Judges were randomly assigned to MDLS, we could use individual judges' propensity to use different procedural devices, such as a Lone Pine order, as an intent to treat (see Helland and Yun, 2023). While this strategy has been effectively used in several other contexts, judges are not randomly assigned in MDLs (Clopton, 2019). The JPML selects both the judge who hears the MDL and the district in which the MDL takes place. Absent some quasi-experiment, we rely on a structural model to at least partially address potential endogeneity in the fractional response model. We follow the two-step control function approach described in Wooldridge (2011), which builds upon the fractional response framework introduced by Papke and Wooldridge (1996). In the first step, the potentially endogenous variable is regressed on the other covariates in the model, and the residuals from this regression are obtained. In the second step, the fractional response model is estimated using a quasi-likelihood based on the Bernoulli distribution, including both the original regressors and the first-stage residuals. The inclusion of the residuals helps control for the endogeneity by proxying for the omitted variables that cause correlation between the regressor and the error term. However, since we have no valid excluded instruments, identification relies on functional form assumptions—that is, on the nonlinearity of the model structure rather than variation from an external instrument. Without an exogenous instrument, the resulting estimates should be interpreted cautiously, as identification is sensitive to specification choices.

One final issue with the fractional response model is the computation of the Average Partial Effects (APE). The APE is the effect of x on y averaged across all cases in the sample. We calculate the APE by determining the partial effect for each case and then averaging these partial effects across all cases in the sample. By contrast, for most models, one would compute the partial effect at the average (PEA) of all other variables, which is the marginal effects with all variables set at their averages. The problem is that PEA is not meaningful for dichotomous variables, as it describes no observations in the sample; hence, we compute APE. Since the variables of interest, Lone Pine orders, and bellwether trials are binary, the APE is the difference in resolution rates between treated periods ($x_{it} = 1$) and untreated periods ($x_{it} = 0$) averaged across all periods. In the logit models used with the FJC data, the APE is easily computable. Wooldridge (2011, 2019) illustrates the average structural function used to compute APE in the fractional response probit model.

*4.3 Results*

Table 3 presents the descriptive statistics for the JPML sample. The Table also breaks out the sample by the period in which a Lone Pine order or bellwether trial process was in effect. Since the JPML data are annual, we treat a year as having a Lone Pine order or bellwether trial process if either case management device was in effect for any portion of the year. The results are robust to treating the orders as starting in the next year.

Table 4 presents the results of the fractional response probit with MDL fixed effects. Column 1 uses all the MDLs in the JPML sample and treats Lone Pine orders and bellwether trial processes as exogenous. The dependent variable is the proportion of the cases resolved in a given year. The impact of both Lone Pine orders and bellwether trial process orders going into effect is positive and significant. In column 2, we present the estimates of the APE. In the case of Lone Pine orders, the impact is 0.3, indicating that the proportion of cases resolved each year a Lone Pine order is in effect more than doubles. For bellwether trial processes, the impact is smaller, 0.15, suggesting a 66% increase in resolutions.

Column 3 of Table 4 presents the results for the full sample of MDLs, where Lone Pine orders or bellwether trial processes are treated as endogenous. The impact of Lone Pine orders is statistically significant, while the impact of bellwether trial processes is no longer significant. When we treat them as endogenous, the APE on Lone Pine orders suggests a three-fold increase in the proportion of cases resolved.

In columns 5 and 7 of Table 4, we restrict the sample to personal injury cases. Burch (2019) finds that personal injury MDLs, particularly those involving pharmaceuticals, differ significantly from other MDLs.[38] Moreover, Lone Pine orders and bellwether trial processes are relatively rare in non-personal injury MDLs. The personal injury MDL results are consistent with the full sample. For both the exogenous and endogenous case management order estimations, Lone Pine orders result in a statistically significant increase in the proportion of cases in the MDL that are resolved in a given year. In both models, the APE is 0.45-0.56, suggesting the proportion of cases resolved in the MDL almost doubles when Lone Pine orders are in effect. Bellwether trial processes increase the proportion of cases resolved.

---

[38] Personal injury MDLs typically involve a larger number of potential plaintiffs, and the universe of potential plaintiffs is generally unknown. Moreover, the harms involved often involve far more complex theories of causation than other MDLs.

The JPML data results suggest that Lone Pine orders at least speed up the resolution of cases within the MDL. Except for defendant fact sheets, which also increase the proportion of cases resolved, we do not find evidence that other case management orders, such as plaintiff fact sheets or profile forms, impact resolution.[39] The estimation also controls for other judicial activity measures, such as rulings on Daubert motions or summary judgment motions, judicial approval of settlements, and the appointment of a special master by the judge.

In Table 5, we decompose the bellwether processes by examining the method used to select trials. As noted above, there are essentially four broad methods of selecting the cases for bellwether trials: plaintiffs nominate trials, defendants nominate trials, courts select trials, and/or cases are randomly chosen. These are not mutually exclusive, and different cases have different elements. One issue is that the number of cases in each cell becomes relatively small—for example, several cases have a pool of trial cases chosen by both defendants and plaintiffs. If both sides submit a list and cases are selected from that list by the judge, we will code the case as plaintiffs, defendants, and the court choosing trials. Ideally, it would be possible to have a control for each of the different combinations. However, there are not enough MDLs in the different combination cells to estimate the model. Table 2 presents the different selection mechanisms.

In column 1 of Table 5, we provide estimates of the different selection mechanisms. For all MDLs, when plaintiffs nominate the trials or cases are randomly selected, the number of cases resolved increases conditional on starting a bellwether process. When we turn to only personal injury MDLs in column 3, plaintiff-nominated trials remain significant, but random selection is not. By contrast, defendants having a role in the nominating process reduces the number of resolutions.

---

[39] The positive and statistically significant association between defendant fact sheet (DFS) orders and resolution rates in the JPML data may reflect a judicial signaling effect rather than the substantive content of the fact sheets themselves. While DFS orders typically impose fewer burdens than plaintiff fact sheets, their issuance may signal that the court expects the parties to begin preparing for more meaningful case activity, such as settlement negotiations, coordinated discovery, or eventual trial scheduling. This signaling aligns with the broader theory of managerial judging in MDLs, in which judges exert informal pressure to move cases toward resolution through procedural cues and structural expectations. A DFS order may therefore act as a proxy for judicial engagement or momentum, particularly in MDLs where other forms of early case management are limited.

One limitation of the JPML dataset is that it does not provide information on how the case was resolved. One possibility is that plaintiffs, unable to comply with the evidentiary burden of Lone Pine orders, may abandon their claims or that judges may dismiss their claims when the plaintiffs fail to produce the required evidence. An alternative possibility is that plaintiffs opt to settle in the face of the Lone Pine order's evidentiary burden. Of course, both effects may occur. In the next section, we turn to the FJC/AO data to attempt to untangle these effects

## 5. Estimation Method and Results: FJC data

### 5.1 Discrete-Time Hazard Model

The FJC data provides case-level information, including the filing date and the date the case was resolved. From this, it is possible to construct the duration of the case. Unlike the JPML data, which provides a count of the number of cases resolved in a given year, the FJC data identifies whether a specific case in the MDL was resolved. We link the data to the dates the case management orders took effect, allowing us to estimate the impact of a judge issuing a Lone Pine order, for example, on the probability that a case will resolve and the type of resolution. Effectively, we are estimating the likelihood that a case resolves in a given time period. Although the data allows us to estimate the probability of joining the MDL each day during the sample period, the time-varying covariates make daily data cumbersome. For this reason, we estimate the model using monthly data, which enables us to estimate the probability that the case is resolved each month after it is filed. For example, if the hypothetical case of Smith versus

Acme Pharmaceutical took 18 months to resolve, we would create 18 observations. If a Lone Pine order went into effect in the 12$^{th}$ month of Smith v. Acme, the indicator variable for Lone Pine orders would equal 1 for all months after 11. If, in month 18, the case is resolved via settlement, the settlement variable would equal one in month 18 and zero otherwise. Table 6 presents the summary statistics for the case-month FJC sample.

To provide a sense of how the estimation works, consider Figure 4, which graphs the impact of the Vioxx Lone Pine Order on resolving cases in the Vioxx MDL. The graph of the number of resolved cases suggests that the Vioxx Lone Pine order had a substantial impact on the total number of resolved cases, consistent with the JPML data results. However, it appears that much of the increase in resolved cases stems from settlements rather than dismissals. For many reasons, however, the Vioxx case may be sui generis. In the next section, we provide evidence on the relative importance of these two effects averaged across the full sample of personal injury MDLs.

We estimate a discrete-time hazard model for the months of filing until resolution, controlling for case management orders, the number of cases, and the number of trials, all of which are captured by time-varying covariates. One concern with the JPML data is that average resolution rates will rise mechanically over time; as the time moves toward the closure of the MDL, the percentage of cases approaches 100%. Even with the JPML data, the annual average resolution rate remains informative, as it reveals systematic variation in resolution trajectories across MDLs and time and whether those are impacted by a Lone Pine order or bellwether process being implemented in that year.[40]

---

[40] For example, in the *Zimmer NexGen Knee Implant* MDL, resolution rates remained sluggish for several years until a Lone Pine order prompted mass dismissals and inventory settlements. In contrast, the *Xarelto* MDL experienced relatively rapid resolutions after a global settlement framework was put in place, with a later Lone Pine order used to clean up straggler claims. The CR Bard Pelvic Mesh MDL similarly saw sharp increases in resolution rates

This concern is addressed directly by the discrete-time hazard model applied to the individual case-level data. Unlike the JPML data, where the dependent variable is the average annual resolution rate for each MDL, the hazard model estimates the conditional probability that a particular case will resolve in a given month, given that it is still pending. This structure accounts for the underlying hazard rate, which naturally rises as MDLs mature, while allowing us to estimate whether specific case management interventions (e.g., Lone Pine orders or bellwether processes) significantly alter the probability of resolution in any given period. The hazard model, by conditioning on survival time and including both MDL fixed effects and calendar time controls, captures whether these tools accelerate the pace of resolution for individual cases.

In all models, we treat cases that have not been resolved by the creation of the dataset as censored, i.e., resolution equals zero for all months. We also treat the handful of cases resolved in trials or remanded by the MDL judge as censored. As noted above, a case can end in one of three ways: dismissal, settlement, or the plaintiff dropping it unilaterally. We first estimate the model with only a single risk. Thus, when estimating the impact of case management orders on settlements, we treat cases that are resolved via a drop or dismissed as censored. The probability of a case ending in one of three outcomes is given by

$$y_{ij} = \alpha_j + \beta Lone\ Pine + \gamma Bellwether + X'_{it}\delta + \mu year + \vartheta \log(t) + \varepsilon_{ij},$$

where the type of resolution $y_{ij} = 1$ if the case resolves via the relevant mechanism and 0 otherwise in any month *t*. Lone Pine and bellwether retain their meanings from above and $X'_{it}$ are

---

following sequential waves of inventory settlements. These patterns are consistent with the theory that procedural tools, such as Lone Pine orders and bellwether processes, do more than formalize inevitable outcomes; they materially shift the timing and clustering of case resolutions in ways that are observable in both aggregate and individual-level data.

control variables described above, and $log(t)$ is the logistic hazard function, which captures the possibility that the likelihood of a case resolving varies over the months it has been in the risk set (i.e., the longer a case has been in the risk set, the higher the risk of resolution).

We estimate the model with and without MDL fixed effects ($\alpha_j$). fixed effects to control for unobserved, time-invariant differences across MDLs, such as judge style, litigation type, or complexity, that may influence resolution timing, thereby reducing omitted variable bias. Comparing both specifications provides insight into whether estimated effects are driven by differences across MDLs or changes within them over time. The same logic applies to the competing risks models: fixed effects require within-MDL variation across resolution types (e.g., settlement vs. dismissal). MDLs in which all cases resolve through a single outcome contribute no identifying variation to the competing hazard model and are excluded when fixed effects are included.

We estimate a discrete-time hazard model with independent risks using the cloglog model; $\Pr(y_{it}|X) = 1 - \exp[-\exp(X\beta)]$, where X are the independent variables discussed above, the three resolution types may interact. To capture this effect, we also estimate a competing risk model, assuming that the discrete interval-specific hazard is multinomial logistic. The multinomial logit can be seen as a proportional odds hazard model, i.e.,

$$h_m(t) = \frac{\exp(X_{mt})}{\sum_{j=1}^{3} \exp(X_{mt})}$$

where $h_m(t)$ denotes the hazard for exit into state $m$ (dropped, dismissed, or settled), and, as above, the hazard function is logistic (i.e., $log(t)$). A case can only exit the MDL via settlement, drop, or dismissal, and we treat not exiting as the base case. Thus, the marginal effects are the impact of exiting via exit type *m* rather than continuing.[41],[42]

*5.2 Results*

Table 7 presents the results for the FJC data estimation without MDL fixed effects, while Table 8 displays the results with MDL fixed effects. We also estimate each model classifying the outcome by only the FJC/AO classifications, the FJC/AO classifications modified by information from the PACER docket for the individual case, and the FJC/AO classifications modified by the SCALES classification in matched cases. Lone Pine orders have a positive and significant impact on the probability that a case is resolved by a plaintiff unilaterally dropping the claim. For each of the different drop classification methods, the impact on drops remains significant, with an APE ranging from 0.007 (FJC) to 0.015 (SCALES). Based on the FJC classification, the probability of a case being dropped by the plaintiff is .006 in any given month, so a .7 percentage point increase would more than double the probability. Given that a sizable fraction of these cases are consistently classified as a unilateral drop in each of the classification mechanisms, this suggests that Lone Pine orders do increase the likelihood that plaintiffs unilaterally abandon their claims in the face of complying with a Lone Pine order. For the FJC and PACER measure of unilateral drops, we find that bellwether trial victories by plaintiffs are associated with a lower probability of a drop in any given month, perhaps suggesting that bellwether victories convey a signal of increased case values in an eventual settlement.

---

[41] Using discrete-time hazard models has a long history in litigation studies, beginning with Kessler (1996). See also Fenn and Rickman (1999, 2013) and Boyd and Hoffman (2012).
[42] Absent a quasi-experiment, we do not have a method for estimating a model that treats Lone Pine orders or bellwether trial processes as endogenous. The two-step process used in the fractional response probit is invalid for discrete probit or logit models.

The impact of Lone Pine orders on the probability a case is resolved by dismissal is not significant for the FJC or PACER classification mechanism. Interestingly, for the SCALES classification, which reclassified the largest proportion of cases, we do find a statistically significant increase in the likelihood of a dismissal, with an API of .009. Although we find no effect of the bellwether process on the probability of a case being dismissed in a given month, we do find that for the FJC and SCALES data, the number of trials completed is associated with an increase in dismissals, although not for the PACER classification. Moreover, we also find a reduction in the monthly likelihood that a case is dismissed when plaintiffs win more cases.

The impact of Lone Pine orders on the monthly probability of settlement is significant for the FJC and SCALES classifications, but not for the PACER reclassification, which may be due to the reduction in sample size. The impact of the FJC and SCALES measure is an API of 0.02, suggesting that, on average, having a Lone Pine order in effect raises the probability of a settlement by approximately two percentage points, which is nearly double the impact on the monthly probability that a case is dropped. The impact is meaningfully large. The APE is 0.02, indicating that the probability a case is settled in any given month in which a Lone Pine order is in effect rises by 151.5% (from 0.0132 to 0.0332). We also find that implementing a plaintiff fact sheet is associated with a reduction in the likelihood of settlement in any month in which the fact sheet is required. Finally, although we do not find an impact of the bellwether process starting, we do find that as more cases are completed, the probability of settlement declines for all three measures. For the FJC and PACER classifications, the number of bellwether trials won by the plaintiff increases the probability of settlement in a given month.

In Table 8, we estimate the model with competing hazards, which allows a case to resolve in any of the three resolution types. The results are nearly identical for Lone Pine orders.

In the competing hazard models, we find a statistically significant increase in the probability that plaintiffs will unilaterally drop or settle their cases when a Lone Pine order is in effect.

In Table 9, we estimate the single hazard model, including fixed effects. The results for Lone Pine orders are similar to those obtained from the model without MDL fixed effects. We find that, according to either the FJC or SCALES outcome measure, the probability that a plaintiff drops their case increases when a Lone Pine order is in effect. When we include MDL fixed effects, the PACER measure of dropped cases becomes insignificant. For settlement, the impact of a Lone Pine order on the probability of a settlement in a given month is statistically significant. The APIs for the FJC and PACER estimates are slightly smaller than those without fixed effects. However, the API for the SCALES reclassification is 0.03, representing an approximately three percentage point increase, or a 167% change in the probability of settlement in the months when a Lone Pine order is in effect.

For the bellwether process, the impacts on drops and dismissals are not consistent across measures when we include fixed effects. However, for the probability of a settlement in a given month, the impact of the bellwether process is statistically significant and positive, with an API of around two percentage points. Plaintiff fact sheets are positive and significant for all but the SCALES classification. In contrast, plaintiff profile forms, which are considered less burdensome, are associated with lower probabilities of settlement during the months they are in effect. Finally, the number of bellwether trials completed is associated with a small but significant reduction in the probability of settlement each month. In contrast, plaintiff wins are associated with the reverse, although neither is significant in the regressions using the SCALES data to reclassify a subset of the cases.

In Table 10, we estimate the competing hazard version of the model with MDL fixed effects. As in the single hazard model, we find that Lone Pine orders increase the likelihood of either unilateral drops by the plaintiff or settlements in the months they are in effect. The magnitudes are similar to the APEs without fixed effects. In the competing hazard model with MDL fixed effects, bellwether trial processes are associated with increases in the probability that the plaintiff and the defendant will resolve the case via settlement each month after the process starts. The impact of the start of the bellwether process on the probability of settlement is mitigated somewhat by the number of trials completed. However, if the plaintiffs win at trial, the probability of settlement increases by the same amount, suggesting that a defendant's win reduces the likelihood of settlement, while a plaintiff's win increases it.

The evidence on plaintiff fact sheets also differs across specifications. Without MDL fixed effects, plaintiff fact sheets appear to increase the likelihood of settlement each month after they are required, but the effect is not significant with MDL fixed effects. By contrast, plaintiff profile forms are associated with a decreased probability of settlement after implementation, which is consistent with the single hazard model for at least two measures of settlement.

Tables 11 and 12 re-estimate the duration model above, but we add controls for how the court selects cases for trial within the bellwether process. We also include the number of bellwether trials completed and the number of trials, if any, won by plaintiffs. As before, we first estimate the models as a single hazard (dropped, dismissed, or settled) and then as a competing hazard model, which captures the interrelationship between the three outcomes. In this specification, we focus solely on the outcomes defined by the FJC/AO codes, although the results are similar for the PACER and SCALES-augmented definitions. In the JPML results above, bellwether processes that involved the plaintiffs nominating cases for trial and/or randomly selecting cases for trial were associated with a higher proportion of cases being resolved (although only the role of plaintiffs in nominating cases is robust to examining only personal injury MDLs). The results from the FJC decomposition regression indicate that increasing the number of bellwether trials completed increases the likelihood of dismissal and decreases the likelihood of a settlement. The impact of a trial is to reduce the probability of a settlement by approximately 0.6 percentage points, compared to bellwether processes that have had no trials, which increase the probability of settlement by 2.2 percentage points each month. However, if the plaintiffs win one of the bellwether trials, the probability of settlement rises by 0.6 percentage points. Because a bellwether trial must be won by either the plaintiff or the defendant, this means that a defendant's win reduces the probability of settlement by 0.6 percentage points. Given a 1.32% probability of settlement each month, the impacts are relatively large. The findings on dismissed and dropped cases reflect a smaller impact but in the opposite direction. We also find that the court selecting the cases for trial is associated with a 1.3 percentage point reduction in the probability of settlement in each month following the initiation of the bellwether process. When bellwether cases are randomly selected, we also find a decrease of 2.6 percentage points in the monthly probability of settlement. The results are similar for the competing hazard model.

In Table 12, for the single-hazard model with fixed effects for MDLs, we find that the court selecting cases is associated with a lower probability of settlement in any month, as indicated by the model without MDL fixed effects. For the estimation with fixed effects, we find that plaintiffs nominating cases and courts selecting cases for trial are associated with a lower likelihood of a judge dismissing the case in any given month. The conclusion is that the earlier finding of bellwether processes promoting settlement remains robust when controls for the selection method are included. However, the evidence on case outcome does not suggest that these different selection mechanisms are driving the settlement result.

One outstanding issue is that the nature and timing of settlement can vary significantly across MDLs. Some MDLs are resolved primarily through inventory settlements, in which defendants negotiate individual deals with plaintiffs' lawyers to settle some portion of their portfolio of cases. Others reach a global settlement, typically negotiated by the plaintiffs' steering committee, that applies to similarly situated plaintiffs in the MDL regardless of who represents them.[43] To account for these dynamics, we include three additional time-varying indicators in our analysis. We code an "inventory settlement" indicator as one in all periods after the first cases are marked as settled in PACER, the FJC dataset, or press reports, in the absence of a publicly announced global settlement. We also include a "global settlement" indicator that turns on in the month when the court docket or press reports indicate that the parties have reached a global settlement.[44]

To capture potential judicial delays that may impede resolution activity, we include a third time-varying indicator for whether the judge stayed the MDL proceedings while the parties were negotiating. In some MDLs, such as Zimmer NexGen Knee Implant (MDL 2272), proceedings were paused by the court to allow parties to negotiate, in this case, unsuccessfully. Such stays may influence the hazard rate by temporarily reducing the probability of resolution.

---

[43] It is worth noting that even global settlements can include a substantial number of cases that do not settle, either because they do not qualify for settlement under the negotiated terms or do not agree to the settlement.

[44] The *Stryker Hip Implant* MDL (MDL 2441) provides an example of this concern. Although the initial global settlement was announced in 2014, a second wave of inventory settlements followed in 2016–2017 for patients not covered by the original agreement. Similarly, in the NFL Concussion MDL (MDL 2323), when the global settlement was initially announced in 2013, a sizable fraction of plaintiffs did not initially agree to the terms. Although the final deadline was August 2017, plaintiffs who missed the deadline were allowed to join for several months afterward. In fact, only 187 plaintiffs out of almost 20,000 did not eventually settle while the MDL was active, and their outcomes appear to be split between those who abandoned their cases and those who ultimately joined the settlement. More broadly, settlement in individual cases does not occur simultaneously across all plaintiffs, even in global resolutions. Timing and eligibility vary based on the injury status, medical review, and administration of the settlement program.

These additional controls enable us to isolate the impact of case management tools (such as Lone Pine orders or bellwether processes) from broader contextual factors that also influence when and how cases are resolved.

We estimate the model using only the single hazard model and the FJC definition of outcomes. We find that the initiation of inventory settlements is associated with a decrease in the likelihood that a case will be dropped in the months following the start of inventory settlements. The APE is .5 percentage points and is consistent with the view that parties are reevaluating the value of cases that might have dropped in light of a defendant's demonstrated willingness to resolve at least some claims. We also find that the judge staying the proceedings during settlement negotiations is, perhaps unsurprisingly, associated with a decline in the probability that a case will be dismissed—which, in some sense, follows from the definition of a stay in proceedings. Finally, we find that in the months following the announcement of a global settlement, the probability that an individual case will settle declines by 1.7 percentage points, consistent with the anecdotal evidence above that the initial announcement of a global settlement does not immediately resolve all cases in the MDL.

However, the impact of Lone Pine orders on settlement and drops is consistent with our earlier results. We find that in periods with a Lone Pine order in effect, the probability a plaintiff unilaterally abandons their claim increases by .6 percentage points, and the probability of a settlement each month rises by 1.8%, almost precisely offsetting the decrease in the likelihood of settlement associated with the announcement of a global settlement.

Finally, in Table 14, we examine the interplay between bellwether processes and Lone Pine orders. Across our sample, several cases have both a bellwether process and a Lone Pine order. In all but one case, Heparin MDL (1953), the Lone Pine order occurs after the bellwether process has started. We find that when the Lone Pine order precedes the start of the Bellwether process, the likelihood of settlement in any given month is reduced, likely because the parties are waiting to see the outcome of the Bellwether trials. We find no effect on the probability that a case is dropped or dismissed when the Lone Pine order precedes the start of the Bellwether process. We find that cases are less likely to be dropped when the Bellwether process precedes the Lone Pine order, again because plaintiffs are likely waiting to see the outcome of Bellwether trials before deciding whether to abandon their claim.

## 6. Conclusions

The use of case management orders in multidistrict litigation (MDLs) has attracted considerable attention from both judges and researchers. Scholars have advanced two hypotheses on the role of a bellwether trial process and Lone Pine. The first is that Lone Pine orders serve as a method for weeding out frivolous cases, or at least increasing the evidentiary burden on plaintiffs, thereby increasing the incentives for plaintiffs to drop their cases. In this framing, Lone Pine orders are essentially summary judgment motions at an early stage of litigation. Bellwether trial processes, by contrast, are designed to provide plaintiffs and defendants with evidence of the value of claims in the MDL, thereby promoting settlement. An alternative hypothesis, advanced by Burch (2019) among others, is that judges use both Lone Pine orders and a bellwether trial process to encourage plaintiffs to settle. Burch (2019) finds evidence that judges utilize both case management orders once the plaintiff's steering committee has reached an agreement with the defendant. In this framing, the importance of both orders is to promote settlement.

While this study cannot resolve which of these hypotheses is correct, and in many ways, they are not mutually exclusive, it does present evidence consistent with Burch and Engstrom's framing while finding little evidence of an early dismissal effect. Specifically, we find that Lone Pine orders are associated with an increase in the fraction of cases in the MDL that are resolved (using data from JPML). Later-stage Lone Pine orders primarily drive this impact on resolutions, and the likelihood of settlement increases when cases are under a Lone Pine order. We do not find systematic evidence that Lone Pine orders are associated with an increase in the judge's probability of dismissal. Muddying the findings in the "dropped" category of the FJC data suggests that the plaintiff either abandoned the case or entered into a settlement without notifying the court. The probability that the plaintiff drops increases in months when a Lone Pine order is in effect. This finding does not rule out the possibility that Lone Pine orders remove frivolous cases or, at the very least, cases in which the plaintiff's injury is difficult to document. The dropped cases may fall into this category. However, it suggests that Burch and Engstrom's hypothesis, that Lone Pine orders push plaintiffs into settlements, is potentially just as important in explaining the increase in resolved cases documented in the JPML data.

Williams, Margaret S. & Tracey E. George, Who Will Manage Complex Civil Litigation? The Decision to Transfer and Consolidate Multidistrict Litigation, 10 *Journal of Empirical Legal Studies* 10:424, (2013).

Willging, Thomas E., and Emery G. Lee III. "From Class Actions to Multidistrict Consolidations: Aggregate Mass-Tort Litigation After Ortiz." University of Kansas Law Review 58 (2009): 775.

Wooldridge, Jeffery M. (2011) "Fractional Response Models with Endogenous Explanatory Variables and Heterogeneity."
https://www.stata.com/meeting/chicago11/materials/chi11_wooldridge.pdf

Wooldridge, Jeffrey M. (2010b) *Econometric analysis of cross section and panel data*. MIT press.

Zimmerman, Adam S. "The Bellwether Settlement." Fordham Law Review 85 (2017): 2275.
45

**Table 1: Summary of Lone Pine Orders Across MDLs**

| MDL Number | Case Name | Start of Lone Pine Order |
|---:|---|---:|
| 1014 | Orthopedic Bone Screw | 1997 |
| 1348 | Rezulin | 2004 |
| 1431 | Baycol | 2004 |
| 1553 | Silica | 2005 |
| 1657 | Vioxx | 2007 |
| 1699 | Bextra and Celebrex Marketing Sales Practices | 2008 |
| 875 | Asbestos (No VI) | 2009 |
| 1742 | Ortho Evra | 2009 |
| 1763 | Human Tissue | 2009 |
| 1845 | ConAgra Peanut Butter | 2010 |
| 1871 | Avandia Marketing Sales Practices | 2010 |
| 1928 | Trasylol | 2010 |
| 1953 | Heparin | 2010 |
| 1789 | Fosamax | 2012 |
| 1842 | Kugel Mesh Hernia Patch | 2012 |
| 1909 | Gadolinium Contrast Dyes | 2013 |
| 1943 | Levaquin | 2013 |
| 2092 | Chantix (Varenicline) | 2013 |
| 2100 | Yasmin and Yaz (Drospirenone) Marketing Sales Practices | 2013 |
| 1964 | NuvaRing | 2014 |
| 2179 | Oil Spill by the Oil Rig Deepwater Horizon | 2014 |
| 2385 | Pradaxa (Dabigatran Etexilate) | 2014 |
| 2575 | Fluidmaster Inc Water Connector Components | 2014 |
| 2272 | Zimmer NexGen Knee Implant | 2016 |
| 2391 | Biomet M2a Hip Implant | 2016 |
| 2428 | Fresenius GranuFlo/NaturaLyte Dialysate | 2017 |
| 2436 | Tylenol (Acetaminophen) Marketing Sales Practices | 2017 |
| 2545 | Testosterone Replacement Therapy | 2018 |
| 2387 | Coloplast Corp Pelvic Support Systems | 2018 |
| 2187 | CR Bard Inc Pelvic Repair Systems | 2018 |
| 2592 | Xarelto (Rivaroxaban) | 2019 |

**Note:** This table reports the MDL number, MDL title, and year of the first known Lone Pine order issued in each multidistrict litigation (MDL). The years reflect the earliest appearance of such an order identified through PACER docket reviews, prior academic sources, or Drug and Device Law blog entries. Some MDLs may have multiple Lone Pine orders; the table lists the first instance.

**Table 2: Summary of Bellwether Trial Processes in MDLs**

| MDL Number | Case Name | Start of Bellwether Process | plaintiff nominated | defense nominated | court selected | randomly selected |
|---|---|---|---|---|---|---|
| 1355 | Propulsid | 2003 | 1 | 1 | | |
| 1431 | Baycol | 2003 | | | 1 | 1 |
| 1657 | Vioxx | 2005 | 1 | 1 | | |
| 1507 | Prempro | 2006 | 1 | 1 | 1 | 1 |
| 1699 | Bextra and Celebrex Marketing Sales Practices | 2006 | | | | 1 |
| 1708 | Guidant Corp Implantable Defibrillators | 2006 | 1 | 1 | | 1 |
| 1358 | Methyl Tertiary Butyl Ether (MTBE) | 2007 | 1 | | | |
| 1535 | Welding Rod | 2007 | 1 | | | 1 |
| 1726 | Medtronic Inc Implantable Defibrillators | 2007 | 1 | 1 | 1 | 1 |
| 1789 | Fosamax | 2007 | | | | 1 |
| 1836 | Mirapex | 2007 | 1 | 1 | | |
| 1629 | Neurontin Marketing Sales Practices | 2007 | 1 | | 1 | |
| 1760 | Aredia and Zometa | 2008 | 1 | 1 | | |
| 1804 | Stand 'n Seal | 2009 | 1 | 1 | 1 | |
| 1873 | FEMA Trailer Formaldehyde | 2009 | 1 | | 1 | |
| 1943 | Levaquin | 2009 | | | 1 | |
| 2004 | Mentor Corp ObTape Transobturator Sling | 2010 | 1 | | | |
| 2100 | Yasmin and Yaz (Drospirenone) Marketing Sales Practices | 2010 | 1 | 1 | 1 | 1 |
| 1953 | Heparin | 2011 | 1 | | | |
| 2092 | Chantix (Varenicline) | 2011 | 1 | 1 | 1 | |
| 2197 | DePuy Orthopaedics Inc ASR Hip Implant | 2012 | 1 | | | |
| 2272 | Zimmer NexGen Knee Implant | 2012 | 1 | 1 | | |
| 2308 | Skechers Toning Shoe | 2012 | 1 | 1 | | |
| 2385 | Pradaxa (Dabigatran Etexilate) | 2012 | 1 | 1 | | |
| 2158 | Zimmer Durom Hip Cup | 2013 | | | 1 | |
| 2244 | DePuy Orthopaedics Inc Pinnacle Hip Implant | 2013 | | | 1 | |
| 2299 | Actos (Pioglitazone) | 2013 | 1 | | 1 | |
| 2327 | Ethicon Inc Pelvic Repair System | 2013 | | | 1 | |
| 2329 | Wright Medical Technology Inc Conserve Hip Implant | 2013 | 1 | 1 | 1 | |
| 2434 | Mirena IUD | 2013 | 1 | 1 | 1 | |
| 2436 | Tylenol (Acetaminophen) Marketing Sales Practices and | 2013 | 1 | 1 | | |
| 2187 | CR Bard Inc Pelvic Repair Systems | 2013 | 1 | 1 | | |
| 2084 | AndroGel (No II) | 2014 | 1 | 1 | | |
| 2326 | Boston Scientific Corp Pelvic Repair System | 2014 | 1 | 1 | 1 | |
| 2543 | General Motors LLC Ignition Switch | 2014 | 1 | 1 | 1 | 1 |
| 2441 | Stryker Rejuvenate and ABG II Hip Implant | 2014 | 1 | | | |
| 2545 | Testosterone Replacement Therapy | 2015 | 1 | 1 | | |
| 2591 | Syngenta AG MIR162 Corn | 2015 | | | 1 | |
| 2592 | Xarelto (Rivaroxaban) | 2015 | 1 | 1 | 1 | 1 |
| 2440 | Cook Medical Inc Pelvic Repair System | 2015 | 1 | 1 | 1 | |
| 2331 | Propecia (Finasteride) | 2016 | 1 | 1 | 1 | |
| 2641 | Bard IVC Filters | 2016 | | | 1 | |
| 2606 | Benicar (Olmesartan) | 2016 | | | | 1 |
| 2642 | Fluoroquinolone | 2017 | | | 1 | 1 |

**Note:** This table lists MDLs in which a bellwether trial process was identified, along with the year in which the process began. Selection mechanisms vary by MDL and may include nominations by plaintiffs, defendants, random selection, or judicial discretion. The year reflects when a formal case management order establishing the bellwether process was entered on the docket. Information was collected from PACER docket sheets, academic sources, and PACER docket reviews.

Table 3: Summary Statistics of JPML MDL Dataset

| | Full Sample | | MDL with Lonepine Orders | | MDL without Lonepine Orders | | MDLs with Belwether trials | | MDLs without Belwether trials | |
|---|---|---|---|---|---|---|---|---|---|---|
| | Mean | S.D. | Mean | S.D. | Mean | S.D. | Mean | S.D. | Mean | S.D. |
| Fraction of Cases in MDL Resolved | 0.2454 | 0.374 | 0.2702 | 0.3381 | 0.2443 | 0.3755 | 0.2157 | 0.313 | 0.247 | 0.3771 |
| Lone Pine Order | 0.0242 | 0.1536 | 0.5969 | 0.4915 | 0 | 0 | 0.1964 | 0.3979 | 0.0147 | 0.1205 |
| Bellwether Trial Process | 0.0421 | 0.2007 | 0.3566 | 0.4799 | 0.0288 | 0.1672 | 0.8097 | 0.3932 | 0 | 0 |
| Plaintiff Fact Sheet | 0.0838 | 0.2771 | 0.7946 | 0.4048 | 0.0538 | 0.2256 | 0.7855 | 0.4111 | 0.0453 | 0.2081 |
| Plaintiff Profile Form | 0.0177 | 0.132 | 0.1085 | 0.3116 | 0.0139 | 0.1171 | 0.1964 | 0.3979 | 0.0079 | 0.0888 |
| Defendant Fact Sheet | 0.0309 | 0.1731 | 0.2752 | 0.4475 | 0.0206 | 0.1421 | 0.3172 | 0.4661 | 0.0152 | 0.1225 |
| MDL is also a Class Action | 0.2796 | 0.4488 | 0.6899 | 0.4634 | 0.2623 | 0.4399 | 0.5378 | 0.4993 | 0.2655 | 0.4416 |
| Judge Ruled on Summary Judgement Motion | 0.0137 | 0.116 | 0.2132 | 0.4103 | 0.0052 | 0.0722 | 0.1903 | 0.3932 | 0.004 | 0.0629 |
| Judge Ruled on a Daubert Motion | 0.0155 | 0.1237 | 0.2287 | 0.4208 | 0.0065 | 0.0806 | 0.2326 | 0.4231 | 0.0036 | 0.0602 |
| Judge Approved Settlement | 0.0097 | 0.0982 | 0.1705 | 0.3768 | 0.0029 | 0.0542 | 0.1269 | 0.3334 | 0.0033 | 0.0574 |
| Number of Cases in MDL | 0.4549 | 3.5989 | 5.5041 | 14.7453 | 0.2418 | 1.7995 | 3.0659 | 5.9284 | 0.3118 | 3.3687 |
| Number of Tried Cases | 0.6 | 6.7621 | 11.6738 | 31.2394 | 0.1328 | 1.1138 | 1.1723 | 2.1879 | 0.5687 | 6.9246 |
| Fraction of Trials on by Plaintiffs | 0.0093 | 0.1967 | 0.0233 | 0.151 | 0.0087 | 0.1983 | 0.0725 | 0.5344 | 0.0058 | 0.158 |
| Total MDLs Managed by Judge | 2.9165 | 1.7843 | 3.1008 | 1.609 | 2.9087 | 1.791 | 4.0091 | 2.931 | 2.8567 | 1.6792 |
| Total Lone Pine Orders Managed by Judge | 0.016 | 0.2641 | -0.1628 | 0.7516 | 0.0235 | 0.218 | 0.0755 | 0.6008 | 0.0127 | 0.2316 |
| Total Bellwether Managed by Judge | 0.0013 | 0.2663 | -0.1589 | 0.619 | 0.008 | 0.2381 | -0.3353 | 0.7858 | 0.0197 | 0.1858 |
| Total MDL Trials Managed by Judge | 0.065 | 1.1109 | 0.0039 | 0.0623 | 0.0675 | 1.1339 | 0.6556 | 3.4035 | 0.0326 | 0.8054 |
| Total Plaintiff Fact Sheets Managed by Judge | 0.2037 | 0.9721 | 0.969 | 1.0053 | 0.1714 | 0.9574 | 1.577 | 2.6359 | 0.1284 | 0.7129 |
| Total Plainitff Profile Forms Managed by Judge | 0.0819 | 0.7524 | 0.2364 | 0.8566 | 0.0754 | 0.7471 | 0.9094 | 2.3363 | 0.0366 | 0.5093 |
| Total Defendant Fact Sheets Managed by Judge | 0.1186 | 0.8996 | 0.4574 | 0.8508 | 0.1043 | 0.8988 | 1.1752 | 2.7435 | 0.0607 | 0.6147 |
| Plaintiff Nominated Trials | 0.0317 | 0.1752 | 0.1899 | 0.393 | 0.025 | 0.1562 | 0.6103 | 0.4884 | 0 | 0 |
| Defendant Nominated Trials | 0.0221 | 0.1471 | 0.1667 | 0.3734 | 0.016 | 0.1256 | 0.426 | 0.4952 | 0 | 0 |
| Court Selected Trials | 0.019 | 0.1365 | 0.1434 | 0.3512 | 0.0137 | 0.1164 | 0.3656 | 0.4823 | 0 | 0 |
| Cases are Randomly Selected | 0.0124 | 0.1107 | 0.1628 | 0.3699 | 0.0061 | 0.0776 | 0.2387 | 0.4269 | 0 | 0 |
| Observations | 6373 | | 258 | | 6115 | | 331 | | 6042 | |

Note: This table presents descriptive statistics for the sample of multidistrict litigations (MDLs) compiled from data provided by the Judicial Panel on Multidistrict Litigation (JPML). The unit of observation is the MDL-year. Lone Pine and bellwether indicators are coded as active for a year if the corresponding case management order was in effect for any portion of that calendar year.

**Table 4: Fraction of Cases Resolved: JPML Data Estimates from Fractional Probit Models**

| VARIABLES | Exogenous Lone Pine | APE | Endogenous Lone Pine & Bellwether | APE:Endogenous Lone Pine & Bellwether | Exogenous Lone Pine: PI Cases | APE: Exogenous Lone Pine PI | Endogenous Lone Pine & Bellwether: PI Cases | APE:Endogenous Lone Pine & Bellwether PI |
|---|---|---|---|---|---|---|---|---|
| Lone Pine Order | **1.15238*** | .3063 | **2.30267** | .9564 | **1.42162*** | .451 | **3.15796*** | .5651 |
|  | **(0.174)** | ( .0589) | **(1.108)** | ( .2375) | **(0.168)** | ( .0604) | **(0.728)** | ( .4285) |
| Bellwether Trial Process | **0.60209*** | .1534 | 0.92075 | .1976 | **0.79332*** | .2517 | **1.62379** | .5236 |
|  | **(0.168)** | ( .0547) | (1.192) | ( .3707) | **(0.201)** | ( .0733) | **(0.757)** | ( .5241) |
| Plaintiff Fact Sheet | 0.32479 | .0836 | 0.23846 | .0667 | 0.32505 | .1031 | 0.15895 | .072 |
|  | (0.227) | ( .0737) | (0.248) | ( .0783) | (0.275) | ( .0957) | (0.311) | ( .0887) |
| Plaintiff Profile Form | -0.24848 | -.1357 | -0.37715 | -.036 | 0.44769 | .142 | 0.18225 | -.0874 |
|  | (0.577) | ( .2601) | (1.015) | ( .3443) | (0.779) | ( .3944) | (1.703) | ( .4949) |
| Defendant Fact Sheet | **1.03670** | .3663 | **1.23971** | .3656 | **1.17179*** | .3718 | **1.24586** | .408 |
|  | **(0.405)** | ( .1715) | **(0.578)** | ( .1704) | **(0.393)** | ( .1736) | **(0.622)** | ( .1829) |
| Observations | 6,373 | | | | 1,677 | | | |

**Note:** All models include indicator variables equal to one for years in which the court certified a class action, ruled on summary judgment or Daubert motions, approved a settlement, or appointed a special master. Models also control for the number of cases in the MDL, the number of cases tried, and include year and MDL fixed effects. Robust standard errors, clustered at the MDL level, are reported in parentheses. Asterisks (*, **, ***) denote statistical significance at the 10%, 5%, and 1% levels, respectively.

**Table 5: Decomposing Bellwether Trial Selection Methods: JPML Data**

| VARIABLES | All MDLs | APE | PI MDLs only | APE |
|---|---|---|---|---|
| Lone Pine Order | **0.99994*** | .2981 | **1.08942*** | .451 |
|  | **(0.175)** | ( .0602) | **(0.211)** | ( .0604) |
| Bellwether Trial Process | -0.66226 | -.1974 | -0.81232 | .2517 |
|  | (0.436) | ( .46) | (0.618) | ( .0733) |
| Plaintiff Nominated Trials | **1.13481*** | .3383 | **1.87940*** | .1031 |
|  | **(0.437)** | ( .488) | **(0.685)** | ( .0957) |
| Defendant Nominated Trials | -0.04173 | -.0124 | **-0.84915** | .142 |
|  | (0.298) | ( .2278) | **(0.360)** | ( .3944) |
| Court Selected Trials | 0.10251 | .0306 | 0.41980 | .3718 |
|  | (0.312) | ( .1456) | (0.356) | ( .1736) |
| Cases are Randomly Selected | **0.81182*** | .242 | 0.88129 | -.0042 |
|  | **(0.314)** | ( .6806) | (0.853) | ( .0105) |
| Observations | 6,373 |  | 1,677 |  |

**Note:** All models include indicator variables equal to one for years in which the MDL had a plaintiff fact sheet order, a plaintiff profile form order, or a defendant fact sheet order; in which the court certified a class action; ruled on summary judgment or Daubert motions; approved a settlement; or appointed a special master. Models also control for the number of cases in the MDL, the number of cases tried, and include year and MDL fixed effects. Robust standard errors, clustered at the MDL level, are reported in parentheses. Asterisks (*, **, ***) denote statistical significance at the 10%, 5%, and 1% levels, respectively.

Table 6: Summary Statistics of FJC Case-Level MDL Dataset

| | Full Sample | | MDL with Lonepine Orders | | MDL without Lone Pine Orders | | MDLs with Belwether trials | | MDLs without Belwether trials | |
|---|---|---|---|---|---|---|---|---|---|---|
| | Mean | S.D. | Mean | S.D. | Mean | S.D. | Mean | S.D. | Mean | S.D. |
| Case is Dropped FJC | 0.0066 | 0.0811 | 0.0182 | 0.1339 | 0.0052 | 0.0719 | 0.0064 | 0.0798 | 0.0069 | 0.0829 |
| Case is Dropped PACER | 0.0084 | 0.0915 | 0.0199 | 0.1396 | 0.007 | 0.0836 | 0.008 | 0.0888 | 0.0091 | 0.0952 |
| Case is Dropped SCALES | 0.0135 | 0.1153 | 0.0251 | 0.1565 | 0.0121 | 0.1091 | 0.0131 | 0.1138 | 0.014 | 0.1175 |
| Case is Dismissed FJC | 0.0011 | 0.0324 | 0.0007 | 0.026 | 0.0011 | 0.0332 | 0.0006 | 0.0239 | 0.0018 | 0.0419 |
| Case is Dismissed PACER | 0.0097 | 0.0979 | 0.0033 | 0.0576 | 0.0105 | 0.1018 | 0.0116 | 0.107 | 0.0069 | 0.083 |
| Case is Dismissed SCALES | 0.009 | 0.0946 | 0.0102 | 0.1002 | 0.0089 | 0.0939 | 0.0084 | 0.0913 | 0.0099 | 0.0992 |
| Case is Settled FJC | 0.0132 | 0.1142 | 0.0518 | 0.2216 | 0.0085 | 0.0918 | 0.0095 | 0.0972 | 0.0186 | 0.1351 |
| Case is Settled PACER | 0.0137 | 0.1161 | 0.0522 | 0.2224 | 0.0089 | 0.0941 | 0.01 | 0.0994 | 0.019 | 0.1365 |
| Case is Settled SCALES | 0.018 | 0.1331 | 0.0579 | 0.2335 | 0.0131 | 0.1139 | 0.0144 | 0.1191 | 0.0233 | 0.1509 |
| Number of Bellwether Trials Completed | 1.857 | 2.9437 | 0.3108 | 0.8175 | 2.0467 | 3.0525 | 2.8203 | 3.2135 | 0.4593 | 1.7194 |
| Number of Bellwether Trials won by plaintiffs | 0.5809 | 1.7984 | 0.0056 | 0.0745 | 0.6515 | 1.8934 | 0.8581 | 2.1562 | 0.1786 | 0.9528 |
| Lone Pine Order | 0.1093 | 0.312 | 1 | 0 | 0 | 0 | 0.0528 | 0.2237 | 0.1912 | 0.3933 |
| Bellwether Trial Process | 0.592 | 0.4915 | 0.2862 | 0.452 | 0.6295 | 0.4829 | 1 | 0 | 0 | 0 |
| Plaintiff Fact Sheet | 0.8047 | 0.3964 | 0.6268 | 0.4837 | 0.8265 | 0.3786 | 0.939 | 0.2393 | 0.6099 | 0.4878 |
| Plaintiff Profile Form | 0.4839 | 0.4997 | 0.392 | 0.4882 | 0.4952 | 0.5 | 0.6034 | 0.4892 | 0.3106 | 0.4627 |
| Defendant Fact Sheet | 0.6289 | 0.4831 | 0.3258 | 0.4687 | 0.6661 | 0.4716 | 0.7656 | 0.4236 | 0.4307 | 0.4952 |
| MDL is also a Class Action | 0.0475 | 0.2128 | 0.0426 | 0.2018 | 0.0481 | 0.2141 | 0.0717 | 0.258 | 0.0124 | 0.1108 |
| Judge Ruled on Summary Judgement Motion | 0.3588 | 0.4796 | 0.4779 | 0.4995 | 0.3441 | 0.4751 | 0.4839 | 0.4997 | 0.1773 | 0.3819 |
| Judge Ruled on a Daubert Motion | 0.2197 | 0.414 | 0.2878 | 0.4527 | 0.2113 | 0.4083 | 0.3573 | 0.4792 | 0.0201 | 0.1404 |
| approve_settle | 0.4076 | 0.4914 | 0.1535 | 0.3605 | 0.4388 | 0.4962 | 0.4598 | 0.4984 | 0.332 | 0.4709 |
| Judge Appointed a Special Master | 0.13 | 0.3364 | 0.1742 | 0.3793 | 0.1246 | 0.3303 | 0.137 | 0.3439 | 0.1199 | 0.3249 |
| Plaintiff Nominated Trials | 0.3663 | 0.4818 | 0.2443 | 0.4297 | 0.3813 | 0.4857 | 0.6188 | 0.4857 | 0 | 0 |
| Defendant Nominated Trials | 0.2968 | 0.4569 | 0.2365 | 0.4249 | 0.3042 | 0.4601 | 0.5014 | 0.5 | 0 | 0 |
| Court Selected Trials | 0.4159 | 0.4929 | 0.0259 | 0.1589 | 0.4637 | 0.4987 | 0.7026 | 0.4571 | 0 | 0 |
| Cases are Randomly Selected | 0.1143 | 0.3182 | 0.0406 | 0.1973 | 0.1233 | 0.3288 | 0.1931 | 0.3947 | 0 | 0 |
| Global Settlement Agreed | 0.1349 | 0.3416 | 0.0026 | 0.0505 | 0.1511 | 0.3582 | 0.1451 | 0.3522 | 0.1201 | 0.325 |
| Inventory Settlements Started | 0.1176 | 0.3221 | 0 | 0 | 0.132 | 0.3385 | 0.1336 | 0.3403 | 0.0943 | 0.2922 |
| Judge stays proceedings while talks are ongoing | 0.0018 | 0.0424 | 0 | 0 | 0.002 | 0.0449 | 0 | 0 | 0.0044 | 0.0663 |
| Observations | 10440262 | | 1141262 | | 9299000 | | 6180180 | | 4260082 | |

**Note:** This table presents summary statistics for individual cases drawn from the Federal Judicial Center (FJC) civil case database, specifically those associated with personal injury multidistrict litigation (MDLs) between 2005 and 2019. The unit of observation is the case-month. Variables include indicators for Lone Pine orders, bellwether trial processes, and other case management orders (e.g., plaintiff fact sheets, profile forms, and defense fact sheets), as well as resolution outcomes (settled, dismissed, or dropped). Only cases with complete disposition information and valid MDL identifiers are included. Resolution variables are constructed using both FJC disposition codes and supplemental information scraped from PACER dockets and the SCALES data.

**Table 7: Discrete Time Hazard Model for Settlement: FJC Single Outcome Model**

| | Single Hazard Model | | | | | |
|---|---|---|---|---|---|---|
| VARIABLES | Dropped FJC | APE | Dropped PACER | APE | Dropped SCALES | APE |
| Lone Pine Order | **1.057*** | **0.007*** | **0.981*** | **0.008*** | **1.114*** | **0.015*** |
| | **(0.260)** | **(0.002)** | **(0.238)** | **(0.002)** | **(0.207)** | **(0.003)** |
| Bellwether Trial Process | -0.054 | -0.000 | -0.117 | -0.001 | 0.162 | 0.002 |
| | (0.408) | (0.003) | (0.317) | (0.003) | (0.199) | (0.003) |
| Plaintiff Fact Sheet | -0.400 | -0.003 | -0.220 | -0.002 | -0.325 | -0.004 |
| | (0.383) | (0.003) | (0.334) | (0.003) | (0.252) | (0.003) |
| Plaintiff Profile Form | -0.290 | -0.002 | -0.231 | -0.002 | **-0.352**** | **-0.005**** |
| | (0.281) | (0.002) | (0.235) | (0.002) | **(0.143)** | **(0.002)** |
| Defendant Fact Sheet | -0.105 | -0.001 | -0.052 | -0.000 | **0.341*** | **0.004*** |
| | (0.350) | (0.002) | (0.276) | (0.002) | **(0.192)** | **(0.003)** |
| Number of Bellwether Trials Completed | -0.020 | -0.000 | -0.008 | -0.000 | 0.037 | 0.000 |
| | (0.065) | (0.000) | (0.064) | (0.001) | (0.063) | (0.001) |
| Number of Bellwether Trials won by plaintiffs | **-0.313**** | **-0.002**** | **-0.160*** | **-0.001*** | -0.116 | -0.002 |
| | **(0.107)** | **(0.001)** | **(0.090)** | **(0.001)** | (0.083) | (0.001) |
| | Dismissed FJC | APE | Dismissed PACER | APE | Dismissed SCALES | APE |
| Lone Pine Order | -1.447 | -0.002 | -0.321 | -0.004 | **0.884**** | **0.009**** |
| | (1.174) | (0.001) | (0.336) | (0.004) | **(0.418)** | **(0.004)** |
| Bellwether Trial Process | -0.823 | -0.001 | -0.002 | -0.000 | -0.130 | -0.001 |
| | (0.721) | (0.001) | (0.405) | (0.005) | (0.305) | (0.003) |
| Plaintiff Fact Sheet | 0.411 | 0.000 | 0.171 | 0.002 | 0.236 | 0.002 |
| | (0.673) | (0.001) | (0.416) | (0.005) | (0.374) | (0.004) |
| Plaintiff Profile Form | -0.019 | -0.000 | 0.019 | 0.000 | **-0.410**** | **-0.004**** |
| | (0.902) | (0.001) | (0.301) | (0.003) | **(0.193)** | **(0.002)** |
| Defendant Fact Sheet | 0.033 | 0.000 | **0.951**** | **0.011**** | **0.538**** | **0.006*** |
| | (0.582) | (0.001) | **(0.336)** | **(0.004)** | **(0.272)** | **(0.003)** |
| Number of Bellwether Trials Completed | **0.327**** | **0.000**** | 0.019 | 0.000 | **0.196**** | **0.002**** |
| | **(0.134)** | **(0.000)** | (0.074) | (0.001) | **(0.097)** | **(0.001)** |
| Number of Bellwether Trials won by plaintiffs | **-0.669**** | **-0.001**** | **-0.200**** | **-0.002**** | **-0.266*** | **-0.003*** |
| | **(0.242)** | **(0.000)** | **(0.094)** | **(0.001)** | **(0.142)** | **(0.002)** |
| | Settled FJC | APE | Settled PACER | APE | Settled SCALES | APE |
| Lone Pine Order | **1.620*** | **0.021*** | 0.472 | 0.005 | **1.516*** | **0.026*** |
| | **(0.307)** | **(0.005)** | (0.517) | (0.006) | **(0.333)** | **(0.006)** |
| Bellwether Trial Process | 0.469 | 0.006 | 0.373 | 0.004 | 0.495 | 0.009 |
| | (0.525) | (0.007) | (0.513) | (0.006) | (0.365) | (0.006) |
| Plaintiff Fact Sheet | **-1.564*** | **-0.020*** | -0.177 | -0.002 | **-1.250*** | **-0.022*** |
| | **(0.405)** | **(0.006)** | (0.518) | (0.006) | **(0.331)** | **(0.006)** |
| Plaintiff Profile Form | 0.448 | 0.006 | 0.333 | 0.004 | 0.333 | 0.006 |
| | (0.377) | (0.005) | (0.319) | (0.003) | (0.255) | (0.005) |
| Defendant Fact Sheet | 0.727 | 0.009 | 0.067 | 0.001 | 0.490 | 0.008 |
| | (0.448) | (0.006) | (0.566) | (0.006) | (0.319) | (0.006) |
| Number of Bellwether Trials Completed | **-0.515*** | **-0.007*** | **-0.331*** | **-0.004*** | **-0.178*** | **-0.003*** |
| | **(0.107)** | **(0.002)** | **(0.091)** | **(0.001)** | **(0.062)** | **(0.001)** |
| Number of Bellwether Trials won by plaintiffs | **0.293**** | **0.004**** | **0.209*** | **0.002*** | 0.047 | 0.001 |
| | **(0.131)** | **(0.002)** | **(0.063)** | **(0.001)** | (0.078) | (0.001) |

**Note:** All models are discrete-time hazard models estimated using a complementary log-log (cloglog) specification, with time measured in months since the filing date. Each model includes indicator variables equal to one for periods after a class action was certified, the judge ruled on summary judgment or Daubert motions, a settlement was judicially approved, or a special master was appointed. Models also control for the number of cases in the MDL, the number of cases tried, and include year fixed effects. The baseline hazard is modeled using the natural logarithm of the number of months since the filing date. Standard errors are clustered at the MDL level and reported in parentheses. Number of observations: 10,439,877 for the FJC and SCALES models and 8,489,646 for the PACER model. Asterisks (*, **, ***) denote statistical significance at the 10%, 5%, and 1% levels, respectively.

**Table 8: Competing Risks Model for Settlement, Dismissal, and Drop: FJC Data**

| | Competing Hazard Model | | | | | |
|---|---|---|---|---|---|---|
| VARIABLES | Dropped FJC | APE | Dismissed FJC | APE | Settled FJC | APE |
| Lone Pine Order | **1.035*** | **0.006*** | -1.035 | -0.001 | **1.708*** | **0.020*** |
| | **(0.258)** | **(0.002)** | (0.904) | (0.001) | **(0.265)** | **(0.004)** |
| Bellwether Trial Process | -0.045 | -0.000 | -0.716 | -0.001 | 0.535 | 0.006 |
| | (0.361) | (0.002) | (0.890) | (0.001) | (0.575) | (0.007) |
| Plaintiff Fact Sheet | -0.307 | -0.002 | 0.686 | 0.001 | -1.216** | -0.015** |
| | (0.404) | (0.003) | (0.772) | (0.001) | (0.485) | (0.006) |
| Plaintiff Profile Form | -0.414 | -0.003 | -0.061 | -0.000 | 0.223 | 0.003 |
| | (0.302) | (0.002) | (0.794) | (0.001) | (0.385) | (0.005) |
| Defendant Fact Sheet | -0.123 | -0.001 | 0.200 | 0.000 | 0.552 | 0.007 |
| | (0.326) | (0.002) | (0.683) | (0.001) | (0.399) | (0.005) |
| Number of Bellwether Trials Completed | -0.055 | -0.000 | **0.283*** | **0.000** | **-0.467*** | **-0.006*** |
| | (0.070) | (0.000) | **(0.165)** | **(0.000)** | **(0.118)** | **(0.002)** |
| Number of Bellwether Trials won by plaintiffs | **-0.272**** | **-0.002**** | **-0.569**** | **-0.001**** | **0.225*** | 0.003 |
| | **(0.119)** | **(0.001)** | **(0.246)** | **(0.000)** | **(0.135)** | (0.002) |
| | Dropped PACER | APE | Dismissed PACER | APE | Settled PACER | APE |
| Lone Pine Order | **1.321*** | **0.007*** | -0.167 | -0.001 | **1.687*** | **0.021*** |
| | **(0.233)** | **(0.001)** | (0.354) | (0.002) | **(0.260)** | **(0.004)** |
| Bellwether Trial Process | 0.358 | 0.002 | 0.430 | 0.003 | 0.529 | 0.006 |
| | (0.328) | (0.002) | (0.416) | (0.003) | (0.547) | (0.007) |
| Plaintiff Fact Sheet | -0.207 | -0.001 | 0.029 | 0.000 | **-1.195**** | **-0.015**** |
| | (0.370) | (0.002) | (0.439) | (0.003) | **(0.465)** | **(0.006)** |
| Plaintiff Profile Form | **-0.820*** | **-0.004*** | -0.485 | -0.003 | 0.209 | 0.003 |
| | **(0.188)** | **(0.001)** | (0.326) | (0.002) | (0.370) | (0.005) |
| Defendant Fact Sheet | -0.290 | -0.002 | **0.628*** | **0.004*** | 0.524 | 0.006 |
| | (0.246) | (0.001) | **(0.345)** | **(0.003)** | (0.395) | (0.005) |
| Number of Bellwether Trials Completed | -0.054 | -0.000 | -0.104 | -0.001 | **-0.451*** | **-0.006*** |
| | (0.071) | (0.000) | (0.090) | (0.001) | **(0.111)** | **(0.002)** |
| Number of Bellwether Trials won by plaintiffs | **-0.245**** | **-0.001**** | -0.025 | -0.000 | **0.215*** | **0.003*** |
| | **(0.123)** | **(0.001)** | (0.123) | (0.001) | **(0.124)** | **(0.002)** |
| | Dropped SCALES | APE | Dismissed SCALES | APE | Settled SCALES | APE |
| Lone Pine Order | **0.851*** | **0.004*** | 0.066 | 0.000 | **1.701*** | **0.028*** |
| | **(0.269)** | **(0.001)** | (0.443) | (0.002) | **(0.258)** | **(0.005)** |
| Bellwether Trial Process | -0.190 | -0.001 | -0.311 | -0.001 | 0.485 | 0.008 |
| | (0.439) | (0.002) | (0.408) | (0.002) | (0.327) | (0.005) |
| Plaintiff Fact Sheet | -0.523 | -0.003 | 0.560 | 0.002 | **-0.990*** | **-0.016*** |
| | (0.451) | (0.002) | (0.509) | (0.002) | **(0.341)** | **(0.006)** |
| Plaintiff Profile Form | -0.454 | -0.002 | **-0.936*** | **-0.003*** | 0.121 | 0.002 |
| | (0.360) | (0.002) | **(0.297)** | **(0.001)** | (0.234) | (0.004) |
| Defendant Fact Sheet | 0.007 | -0.000 | -0.117 | -0.000 | 0.429 | 0.007 |
| | (0.347) | (0.002) | (0.367) | (0.001) | (0.269) | (0.004) |
| Number of Bellwether Trials Completed | -0.129 | -0.001 | 0.064 | 0.000 | **-0.212*** | **-0.003*** |
| | (0.097) | (0.001) | (0.095) | (0.000) | **(0.066)** | **(0.001)** |
| Number of Bellwether Trials won by plaintiffs | -0.257 | -0.001 | -0.224* | -0.001* | 0.096 | 0.002 |
| | (0.173) | (0.001) | (0.121) | (0.000) | (0.083) | (0.001) |

**Note:** All models are discrete-time hazard models estimated using a multinomial logit specification, with time measured in months since the filing date. Each model includes indicator variables equal to one for periods after a class action was certified, the judge ruled on summary judgment or Daubert motions, a settlement was judicially approved, or a special master was appointed. Models also control for the number of cases in the MDL, the number of cases tried, and include year fixed effects. The baseline hazard is modeled using the natural logarithm of the number of months since filing. Standard errors are clustered at the MDL level and reported in parentheses. Number of observations: 10,439,877 for the FJC and SCALES models and 8,489,646 for the PACER model. Asterisks (*, **, ***) denote statistical significance at the 10%, 5%, and 1% levels, respectively.

Table 9: MDL Fixed Effects Discrete Time Hazard Model: FJC Single Outcome Model

| VARIABLES | Single Hazard Model | | | | | |
|---|---|---|---|---|---|---|
| | Dropped FJC | APE | Dropped PACER | APE | Dropped SCALES | APE |
| Lone Pine Order | **0.860*** | **0.006**** | 0.686 | 0.006 | **1.367**** | **0.018**** |
| | **(0.439)** | **(0.003)** | (0.456) | (0.004) | **(0.568)** | **(0.007)** |
| Bellwether Trial Process | -0.343 | -0.002 | -0.111 | -0.001 | -0.429 | -0.006 |
| | (0.412) | (0.003) | (0.322) | (0.003) | (0.327) | (0.004) |
| Plaintiff Fact Sheet | 0.086 | 0.001 | 0.590 | 0.005 | -0.037 | -0.000 |
| | (0.384) | (0.003) | (0.364) | (0.003) | (0.588) | (0.008) |
| Plaintiff Profile Form | 0.261 | 0.002 | 0.279 | 0.002 | 0.176 | 0.002 |
| | (0.576) | (0.004) | (0.533) | (0.004) | (0.590) | (0.008) |
| Defendant Fact Sheet | 0.143 | 0.001 | 0.449 | 0.004 | -1.093 | -0.014 |
| | (0.482) | (0.003) | (0.530) | (0.004) | (0.690) | (0.009) |
| Number of Bellwether Trials Completed | **-0.098**** | **-0.001**** | -0.019 | -0.000 | 0.069 | 0.001 |
| | **(0.038)** | **(0.000)** | (0.046) | (0.000) | (0.059) | (0.001) |
| Number of Bellwether Trials won by plaintiffs | **0.115**** | **0.001**** | 0.087 | 0.001 | **0.138*** | **0.002*** |
| | **(0.053)** | **(0.000)** | (0.059) | (0.000) | **(0.082)** | **(0.001)** |
| | Dismissed FJC | APE | Dismissed PACER | APE | Dismissed SCALES | APE |
| Lone Pine Order | 0.796 | 0.001 | -0.170 | -0.002 | **1.706**** | **0.014**** |
| | (0.847) | (0.001) | (0.770) | (0.007) | **(0.741)** | **(0.006)** |
| Bellwether Trial Process | 1.231 | 0.001 | 0.552 | 0.005 | 0.364 | 0.003 |
| | (0.864) | (0.001) | (0.688) | (0.006) | (0.555) | (0.004) |
| Plaintiff Fact Sheet | -0.191 | -0.000 | **1.739**** | **0.016**** | -0.489 | -0.004 |
| | (0.931) | (0.001) | **(0.871)** | **(0.008)** | (0.901) | (0.007) |
| Plaintiff Profile Form | 0.795 | 0.001 | 0.151 | 0.001 | 0.854 | 0.007 |
| | (1.160) | (0.001) | (0.665) | (0.006) | (0.964) | (0.008) |
| Defendant Fact Sheet | 0.799 | 0.001 | 0.753 | 0.007 | 0.276 | 0.002 |
| | (0.589) | (0.001) | (0.585) | (0.005) | (1.286) | (0.010) |
| Number of Bellwether Trials Completed | 0.248 | 0.000 | -0.045 | -0.000 | 0.383 | 0.003 |
| | (0.169) | (0.000) | (0.091) | (0.001) | (0.239) | (0.002) |
| Number of Bellwether Trials won by plaintiffs | -0.090 | -0.000 | 0.070 | 0.001 | -0.150 | -0.001 |
| | (0.184) | (0.000) | (0.118) | (0.001) | (0.308) | (0.002) |
| | Settled FJC | APE | Settled PACER | APE | Settled SCALES | APE |
| Lone Pine Order | **1.153**** | **0.015**** | **1.094**** | **0.014**** | **1.749****** | **0.030****** |
| | **(0.480)** | **(0.006)** | **(0.467)** | **(0.006)** | **(0.628)** | **(0.011)** |
| Bellwether Trial Process | **1.996****** | **0.025****** | **1.909****** | **0.025****** | **1.257**** | **0.022**** |
| | **(0.619)** | **(0.007)** | **(0.589)** | **(0.007)** | **(0.531)** | **(0.009)** |
| Plaintiff Fact Sheet | **1.562****** | **0.020****** | **1.414****** | **0.018****** | 0.956 | 0.016 |
| | **(0.527)** | **(0.007)** | **(0.493)** | **(0.006)** | (0.748) | (0.013) |
| Plaintiff Profile Form | **-1.081**** | **-0.014**** | **-1.045**** | **-0.014**** | -0.507 | -0.009 |
| | **(0.457)** | **(0.006)** | **(0.448)** | **(0.006)** | (0.709) | (0.012) |
| Defendant Fact Sheet | -0.133 | -0.002 | 0.043 | 0.001 | -0.847 | -0.014 |
| | (0.443) | (0.006) | (0.370) | (0.005) | (0.780) | (0.013) |
| Number of Bellwether Trials Completed | **-0.424****** | **-0.005****** | **-0.396****** | **-0.005****** | -0.091 | -0.002 |
| | **(0.106)** | **(0.001)** | **(0.100)** | **(0.001)** | (0.081) | (0.001) |
| Number of Bellwether Trials won by plaintiffs | **0.291****** | **0.004****** | **0.270****** | **0.004****** | 0.068 | 0.001 |
| | **(0.105)** | **(0.001)** | **(0.096)** | **(0.001)** | (0.075) | (0.001) |

Note: All models are discrete-time hazard models estimated using a complementary log-log (cloglog) specification, with time measured in months since the filing date. Each model includes indicator variables equal to one for periods after a class action was certified, the judge ruled on summary judgment or Daubert motions, a settlement was judicially approved, or a special master was appointed. Models also control for the number of cases in the MDL, the number of cases tried, and include year and MDL fixed effects. The baseline hazard is modeled using the natural logarithm of the number of months since the filing date. Standard errors are clustered at the MDL level and reported in parentheses. Number of observations: 10,439,877 for the FJC and SCALES models and 8,489,646 for the PACER model. Asterisks (*, **, ***) denote statistical significance at the 10%, 5%, and 1% levels, respectively.

**Table 10: MDL Fixed Effects Competing Risks Model: Settlement, Dismissal, and Drop: FJC Data**

| | Competing Hazard Model | | | | | |
|---|---|---|---|---|---|---|
| VARIABLES | Dropped FJC | APE | Dismissed FJC | APE | Settled FJC | APE |
| Lone Pine Order | **1.035**** | **0.006**** | 0.982 | 0.001 | **1.046**** | **0.013**** |
| | **(0.514)** | **(0.003)** | (0.899) | (0.001) | **(0.516)** | **(0.006)** |
| Bellwether Trial Process | -0.007 | -0.001 | 1.297 | 0.001 | **2.099**** | **0.026**** |
| | (0.424) | (0.003) | (0.886) | (0.001) | **(0.704)** | **(0.008)** |
| Plaintiff Fact Sheet | -0.129 | -0.001 | -0.531 | -0.001 | 0.889 | 0.011 |
| | (0.499) | (0.003) | (0.970) | (0.001) | (0.713) | (0.009) |
| Plaintiff Profile Form | 0.521 | 0.004 | 1.133 | 0.001 | **-1.260**** | **-0.016**** |
| | (0.584) | (0.004) | (1.299) | (0.001) | **(0.459)** | **(0.005)** |
| Defendant Fact Sheet | 0.583 | 0.004 | 0.652 | 0.001 | 0.093 | 0.001 |
| | (0.364) | (0.002) | (0.620) | (0.001) | (0.548) | (0.007) |
| Number of Bellwether Trials Completed | **-0.127**** | **-0.001**** | 0.252 | 0.000 | **-0.449**** | **-0.006**** |
| | **(0.048)** | **(0.000)** | (0.169) | (0.000) | **(0.120)** | **(0.001)** |
| Number of Bellwether Trials won by plaintiffs | **0.124**** | **0.001**** | -0.105 | -0.000 | **0.312**** | **0.004**** |
| | **(0.056)** | **(0.000)** | (0.180) | (0.000) | **(0.112)** | **(0.001)** |
| | Dropped PACER | APE | Dismissed PACER | APE | Settled PACER | APE |
| Lone Pine Order | **1.249**** | **0.006**** | 0.416 | 0.002 | **1.033**** | **0.013**** |
| | **(0.503)** | **(0.003)** | (0.269) | (0.002) | **(0.514)** | **(0.006)** |
| Bellwether Trial Process | -0.007 | -0.001 | 0.499 | 0.003 | **2.089**** | **0.027**** |
| | (0.346) | (0.002) | (0.994) | (0.006) | **(0.661)** | **(0.008)** |
| Plaintiff Fact Sheet | -0.169 | -0.001 | 0.759 | 0.005 | 0.733 | 0.009 |
| | (0.453) | (0.002) | (0.637) | (0.004) | (0.681) | (0.009) |
| Plaintiff Profile Form | 0.265 | 0.002 | 0.309 | 0.002 | **-1.224**** | **-0.016**** |
| | (0.703) | (0.004) | (0.794) | (0.005) | **(0.452)** | **(0.006)** |
| Defendant Fact Sheet | 0.600 | 0.003 | 0.972 | 0.006 | 0.285 | 0.003 |
| | (0.374) | (0.002) | (0.886) | (0.006) | (0.481) | (0.006) |
| Number of Bellwether Trials Completed | **-0.131**** | **-0.001**** | -0.027 | -0.000 | **-0.423**** | **-0.005**** |
| | **(0.049)** | **(0.000)** | (0.098) | (0.001) | **(0.114)** | **(0.001)** |
| Number of Bellwether Trials won by plaintiffs | **0.125**** | **0.001*** | 0.044 | 0.000 | **0.291**** | **0.004**** |
| | **(0.059)** | **(0.000)** | (0.141) | (0.001) | **(0.102)** | **(0.001)** |
| | Dropped SCALES | APE | Dismissed SCALES | APE | Settled SCALES | APE |
| Lone Pine Order | **1.936**** | 0.010*** | 0.567 | 0.002 | **1.315**** | **0.021**** |
| | **(0.467)** | (0.002) | (0.447) | (0.001) | **(0.575)** | **(0.009)** |
| Bellwether Trial Process | **-0.769*** | **-0.004*** | -0.536 | -0.002 | 0.543 | 0.009 |
| | **(0.417)** | **(0.002)** | (0.597) | (0.002) | (0.400) | (0.006) |
| Plaintiff Fact Sheet | -0.079 | -0.000 | -0.083 | -0.000 | -0.578 | -0.010 |
| | (0.560) | (0.003) | (0.519) | (0.002) | (0.635) | (0.010) |
| Plaintiff Profile Form | 0.779* | 0.004* | -0.664 | -0.002 | 0.166 | 0.003 |
| | (0.428) | (0.002) | (0.498) | (0.002) | (0.707) | (0.012) |
| Defendant Fact Sheet | 0.413 | 0.002 | 0.149 | 0.000 | 0.413 | 0.007 |
| | (0.550) | (0.003) | (0.577) | (0.002) | (0.562) | (0.009) |
| Number of Bellwether Trials Completed | **-0.228**** | **-0.001**** | 0.221 | 0.001 | -0.114 | -0.002 |
| | **(0.088)** | **(0.000)** | (0.141) | (0.000) | (0.083) | (0.001) |
| Number of Bellwether Trials won by plaintiffs | 0.052 | 0.000 | -0.197 | -0.001 | 0.089 | 0.001 |
| | (0.199) | (0.001) | (0.169) | (0.001) | (0.074) | (0.001) |

**N**ote: All models are discrete-time hazard models estimated using a multinomial logit specification, with time measured in months since the filing date. Each model includes indicator variables equal to one for periods after a class action was certified, the judge ruled on summary judgment or Daubert motions, a settlement was judicially approved, or a special master was appointed. Models also control for the number of cases in the MDL, the number of cases tried, and include year and MDL fixed effects. The baseline hazard is modeled using the natural logarithm of the number of months since the filing date. Standard errors are clustered at the MDL level and reported in parentheses. Number of observations: 10,439,877 for the FJC and SCALES models and 8,489,646 for the PACER model. Asterisks (*, **, ***) denote statistical significance at the 10%, 5%, and 1% levels, respectively.

Table 11: Bellwether Trial Process Decomposition by Selection Method: FJC Data

| VARIABLES | Single Hazard Model | | | | | | Competing Hazard Model | | | | | |
|---|---|---|---|---|---|---|---|---|---|---|---|---|
| | Dropped | APE | Dismissed | APE | Settled | APE | Dropped | APE | Dimissed | APE | Settled | APE |
| Lone Pine Order | **1.094*** ** | **0.007*** ** | -1.384 | -0.001 | **1.379*** ** | **0.017*** ** | **1.172*** ** | **0.007*** ** | -1.357 | -0.001 | **1.414*** ** | **0.017*** ** |
| | **(0.248)** | **(0.002)** | (1.269) | (0.001) | **(0.285)** | **(0.004)** | **(0.267)** | **(0.002)** | (1.274) | (0.001) | **(0.303)** | **(0.004)** |
| Bellwether Trial Process | 0.169 | 0.001 | -2.707* | -0.003* | **1.773*** ** | **0.022*** ** | 0.213 | 0.001 | **-2.700*** | **-0.003*** | **1.819*** ** | **0.022*** ** |
| | (0.591) | (0.004) | (1.387) | (0.002) | **(0.655)** | **(0.008)** | (0.604) | (0.004) | **(1.389)** | **(0.002)** | **(0.671)** | **(0.008)** |
| Number of Bellwether Trials Completed | -0.039 | -0.000 | **0.366*** ** | **0.000*** ** | **-0.632*** ** | **-0.008*** ** | -0.051 | -0.000 | **0.366*** ** | **0.000*** ** | **-0.639*** ** | **-0.008*** ** |
| | (0.056) | (0.000) | **(0.122)** | **(0.000)** | **(0.083)** | **(0.001)** | (0.059) | (0.000) | **(0.124)** | **(0.000)** | **(0.088)** | **(0.001)** |
| Number of Bellwether Trials won by plaintiffs | **-0.298**** | **-0.002**** | **-0.684**** | **-0.001**** | **0.469*** ** | **0.006*** ** | **-0.287**** | **-0.002**** | **-0.687**** | **-0.001**** | **0.477*** ** | **0.006*** ** |
| | **(0.121)** | **(0.001)** | **(0.309)** | **(0.000)** | **(0.102)** | **(0.001)** | **(0.121)** | **(0.001)** | **(0.312)** | **(0.000)** | **(0.101)** | **(0.001)** |
| Plaintiff Nominated Trials | -0.066 | -0.000 | **2.837*** | 0.003 | 0.073 | 0.001 | -0.063 | -0.000 | **2.832*** | 0.003 | 0.054 | 0.001 |
| | (0.563) | (0.004) | **(1.610)** | (0.002) | (0.851) | (0.011) | (0.567) | (0.004) | **(1.610)** | (0.002) | (0.865) | (0.010) |
| Defendant Nominated Trials | -0.650 | -0.004 | -1.171 | -0.001 | -0.770 | -0.010 | -0.652 | -0.004 | -1.178 | -0.001 | -0.779 | -0.009 |
| | (0.642) | (0.004) | (1.253) | (0.001) | (0.711) | (0.009) | (0.649) | (0.004) | (1.254) | (0.001) | (0.731) | (0.009) |
| Court Selected Trials | 0.137 | 0.001 | 1.246 | 0.001 | **-1.055*** | **-0.013*** | 0.126 | 0.001 | 1.237 | 0.001 | **-1.072*** | **-0.013*** |
| | (0.372) | (0.002) | (0.840) | (0.001) | **(0.553)** | **(0.007)** | (0.379) | (0.003) | (0.851) | (0.001) | **(0.567)** | **(0.007)** |
| Cases are Randomly Selected | 0.335 | 0.002 | -1.259 | -0.001 | **-2.036*** ** | **-0.026*** ** | 0.306 | 0.002 | -1.260 | -0.001 | **-2.047*** ** | **-0.025*** ** |
| | (0.525) | (0.003) | (1.627) | (0.002) | **(0.730)** | **(0.009)** | (0.537) | (0.003) | (1.624) | (0.002) | **(0.747)** | **(0.009)** |

**N**ote: All models are discrete-time hazard models estimated using a complementary log-log (cloglog) or multinomial logit specification, with time measured in months since the filing date. Each model includes indicator variables equal to one for periods after a class action was certified, the judge ruled on summary judgment or Daubert motions, a settlement was judicially approved, or a special master was appointed. Models also control for the number of cases in the MDL, the number of cases tried, and include year fixed effects. The baseline hazard is modeled using the natural logarithm of the number of months since the filing date. Standard errors are clustered at the MDL level and reported in parentheses. Number of observations: 10,439,877 for the FJC and SCALES models and 8,489,646 for the PACER model. Asterisks (*, **, ***) denote statistical significance at the

Table 12: MDL Fixed Effects Model: Bellwether Trial Selection Mechanism Effects: FJC Data

|  | Single Hazard Model | | | | | | Competing Hazard Model | | | | | |
| --- | --- | --- | --- | --- | --- | --- | --- | --- | --- | --- | --- | --- |
| VARIABLES | Dropped FJC | APE | Dismissed FJC | APE | Settled FJC | APE | Dropped FJC | APE | Dimissed FJC | APE | Settled FJC | APE |
| Lone Pine Order | **0.817*** | **0.005*** | 0.633 | 0.001 | **1.151**** | **0.015**** | **1.000*** | **0.006*** | 0.748 | 0.001 | **1.046**** | **0.013**** |
|  | **(0.449)** | **(0.003)** | (0.799) | (0.001) | **(0.479)** | **(0.006)** | **(0.524)** | **(0.003)** | (0.823) | (0.001) | **(0.516)** | **(0.006)** |
| Bellwether Trial Process | 0.116 | 0.001 | 9.607*** | 0.011*** | 2.667* | 0.034* | 0.755 | 0.004 | 9.706*** | 0.010*** | 3.540** | 0.044** |
|  | (2.397) | (0.016) | **(2.816)** | **(0.003)** | **(1.506)** | **(0.018)** | (2.433) | (0.015) | **(2.853)** | **(0.003)** | **(1.770)** | **(0.021)** |
| Number of Bellwether Trials Completed | -0.103*** | -0.001*** | 0.240 | 0.000 | -0.424*** | -0.005*** | -0.130*** | -0.001*** | 0.235 | 0.000 | -0.449*** | -0.006*** |
|  | **(0.037)** | **(0.000)** | (0.177) | (0.000) | **(0.106)** | **(0.001)** | **(0.048)** | **(0.000)** | (0.182) | (0.000) | **(0.120)** | **(0.001)** |
| Number of Bellwether Trials won by plaintiffs | 0.119** | 0.001** | -0.059 | -0.000 | 0.291*** | 0.004*** | 0.127** | 0.001** | -0.069 | -0.000 | 0.312*** | 0.004*** |
|  | **(0.052)** | **(0.000)** | (0.195) | (0.000) | **(0.105)** | **(0.001)** | **(0.055)** | **(0.000)** | (0.194) | (0.000) | **(0.112)** | **(0.001)** |
| Plaintiff Nominated Trials | 0.195 | 0.001 | -8.706*** | -0.010*** | -0.598 | -0.008 | -0.203 | -0.001 | -8.230*** | -0.008*** | -0.406 | -0.005 |
|  | (2.415) | (0.016) | **(2.067)** | **(0.002)** | (0.983) | (0.012) | (2.440) | (0.015) | **(2.045)** | **(0.002)** | (1.070) | (0.013) |
| Defendant Nominated Trials | 1.431 | 0.009 | 8.041*** | 0.009*** | 1.321** | 0.017** | 1.359 | 0.008 | 7.340*** | 0.007*** | 0.791 | 0.009 |
|  | (2.282) | (0.015) | **(2.171)** | **(0.002)** | **(0.640)** | **(0.008)** | (2.366) | (0.015) | **(2.127)** | **(0.002)** | (1.000) | (0.012) |
| Court Selected Trials | -1.667 | -0.011 | -7.479*** | -0.008*** | -1.692* | -0.021* | -1.647 | -0.010 | -7.341*** | -0.007*** | -2.320* | -0.028* |
|  | (2.260) | (0.015) | **(2.430)** | **(0.003)** | **(0.912)** | **(0.011)** | (2.323) | (0.015) | (2.532) | (0.003) | **(1.246)** | **(0.015)** |
| Cases are Randomly Selected | -0.337 | -0.002 | -0.823 | -0.001 | 1.050 | 0.013 | -0.593 | -0.004 | -0.874 | -0.001 | 0.638 | 0.008 |
|  | (0.726) | (0.005) | (1.554) | (0.002) | (1.472) | (0.019) | (0.812) | (0.005) | (1.500) | (0.002) | (1.619) | (0.020) |
| Observations | 10,339,185 | | | | | | 9,421,457 | | | | | |

Note: All models are discrete-time hazard models estimated using a complementary log-log (cloglog) specification, with time measured in months since the filing date. Each model includes indicator variables equal to one for periods after a class action was certified, the judge ruled on summary judgment or Daubert motions, a settlement was judicially approved, or a special master was appointed. Models also control for the number of cases in the MDL, the number of cases tried, and include year fixed effects. The baseline hazard is modeled using the natural logarithm of the number of months since the filing date. Standard errors are clustered at the MDL level and reported in parentheses. Number of observations: 10,439,877. Asterisks (*, **, ***) denote statistical significance at the 10%, 5%, and 1% levels, respectively.

**Table 13: Impact of Global vs. Inventory Settlements on Case Resolution Timing: FJC data**

|  | Single Hazard Model | | | | | |
|---|---|---|---|---|---|---|
| VARIABLES | Dropped FJC | APE | Dismissed FJC | APE | Settled FJC | APE |
| Lone Pine Order | **0.933*** | **0.006*** | -1.476 | -0.002 | **1.451*** | **0.018*** |
|  | **(0.259)** | **(0.002)** | (1.184) | (0.001) | **(0.289)** | **(0.004)** |
| Bellwether Trial Process | -0.016 | -0.000 | -0.843 | -0.001 | 0.635 | 0.008 |
|  | (0.408) | (0.003) | (0.711) | (0.001) | (0.410) | (0.005) |
| Plaintiff Fact Sheet | -0.455 | -0.003 | 0.418 | 0.000 | **-1.885*** | **-0.024*** |
|  | (0.386) | (0.003) | (0.670) | (0.001) | **(0.400)** | **(0.005)** |
| Plaintiff Profile Form | -0.175 | -0.001 | -0.038 | -0.000 | **0.598*** | **0.008*** |
|  | (0.266) | (0.002) | (0.911) | (0.001) | **(0.307)** | **(0.004)** |
| Defendant Fact Sheet | -0.089 | -0.001 | -0.009 | -0.000 | 0.687 | 0.009 |
|  | (0.349) | (0.002) | (0.568) | (0.001) | (0.446) | (0.006) |
| Number of Bellwether Trials Completed | -0.063 | -0.000 | **0.331**** | **0.000**** | **-0.700*** | **-0.009*** |
|  | (0.066) | (0.000) | **(0.132)** | **(0.000)** | **(0.083)** | **(0.001)** |
| Number of Bellwether Trials won by plaintiffs | -0.209 | -0.001 | **-0.683*** | **-0.001**** | **0.628*** | **0.008*** |
|  | (0.137) | (0.001) | **(0.232)** | **(0.000)** | **(0.098)** | **(0.001)** |
| Global Settlement Agreed | -0.431 | -0.003 | -0.405 | -0.000 | **-1.335*** | **-0.017*** |
|  | (0.445) | (0.003) | (0.575) | (0.001) | **(0.417)** | **(0.005)** |
| Inventory Settlements Started | **-1.068**** | **-0.007**** | 0.741 | 0.001 | -0.561 | -0.007 |
|  | **(0.509)** | **(0.003)** | (0.825) | (0.001) | (0.459) | (0.006) |
| Judge stays proceedings while talks are ongoing | -0.392 | -0.003 | **-4.508*** | **-0.005*** | -0.495 | -0.006 |
|  | (0.827) | (0.005) | **(1.320)** | **(0.002)** | (0.597) | (0.007) |

**N**ote: All models are discrete-time hazard models estimated using a complementary log-log (cloglog) specification, with time measured in months since the filing date. Each model includes indicator variables equal to one for periods after a class action was certified, the judge ruled on summary judgment or Daubert motions, a settlement was judicially approved, or a special master was appointed. Models also control for the number of cases in the MDL, the number of cases tried, and include year fixed effects. The baseline hazard is modeled using the natural logarithm of the number of months since the filing date. Standard errors are clustered at the MDL level and reported in parentheses. Number of observations: 10,439,877. Asterisks (*, **, ***) denote statistical significance at the 10%, 5%, and 1% levels, respectively.

**Table 14: Timing-Based Effects of Lone Pine and Bellwether Processes on Resolution Rates: FJC Data**

|  | Single Hazard Model | | | | | |
|---|---|---|---|---|---|---|
| VARIABLES | Dropped FJC | APE | Dismissed FJC | APE | Settled FJC | APE |
| Lone Pine Order | **1.417*** | **0.009*** | -1.428 | -0.001 | **1.656*** | **0.021*** |
|  | **(0.305)** | **(0.002)** | (1.255) | (0.001) | **(0.379)** | **(0.005)** |
| Bellwether Trial Process | 0.195 | 0.001 | -0.816 | -0.001 | 0.481 | 0.006 |
|  | (0.397) | (0.003) | (0.731) | (0.001) | (0.539) | (0.007) |
| Lone Pine Order Precedes Bellwether Process | 0.033 | 0.000 | 1.485 | 0.002 | **-2.000*** | **-0.025*** |
|  | (0.640) | (0.004) | (1.548) | (0.002) | **(0.578)** | **(0.008)** |
| Bellwether Process Precedes Lone Pine Order | **-1.173*** | **-0.008*** | -1.359 | -0.001 | -0.141 | -0.002 |
|  | **(0.603)** | **(0.004)** | (1.211) | (0.001) | (0.688) | (0.009) |

Note: All models are discrete-time hazard models estimated using a complementary log-log (cloglog) specification, with time measured in months since the filing date. Each model includes indicator variables equal to one for periods after a class action was certified, the judge ruled on summary judgment or Daubert motions, a settlement was judicially approved, or a special master was appointed. Models also control for the number of cases in the MDL, the number of cases tried, and include year fixed effects. The baseline hazard is modeled using the natural logarithm of the number of months since the filing date. Standard errors are clustered at the MDL level and reported in parentheses. Number of observations: 10,439,877. Asterisks (*, **, ***) denote statistical significance at the 10%, 5%, and 1% levels, respectively.

**Figure 1: The Number of Cases in the MDL System by Year Resolved**

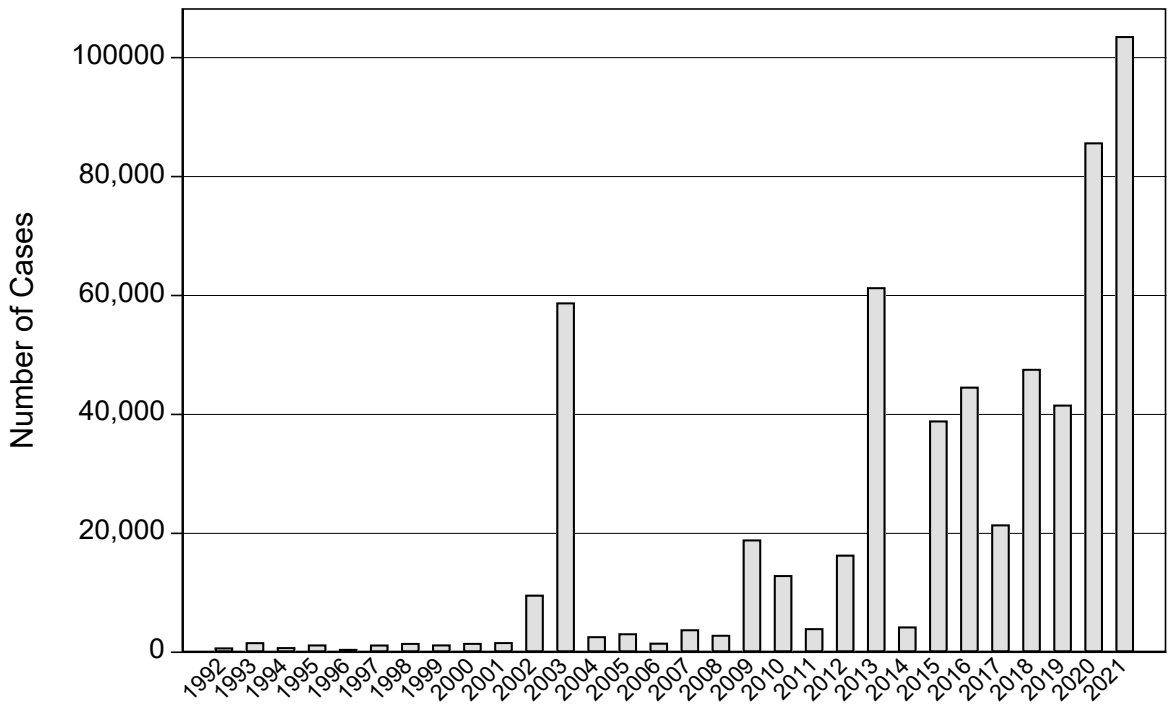

Source: JPML

**Figure 2: Shankley Graph showing the change in outcome classification from FJC codes to PACER Events to SCALES classification**

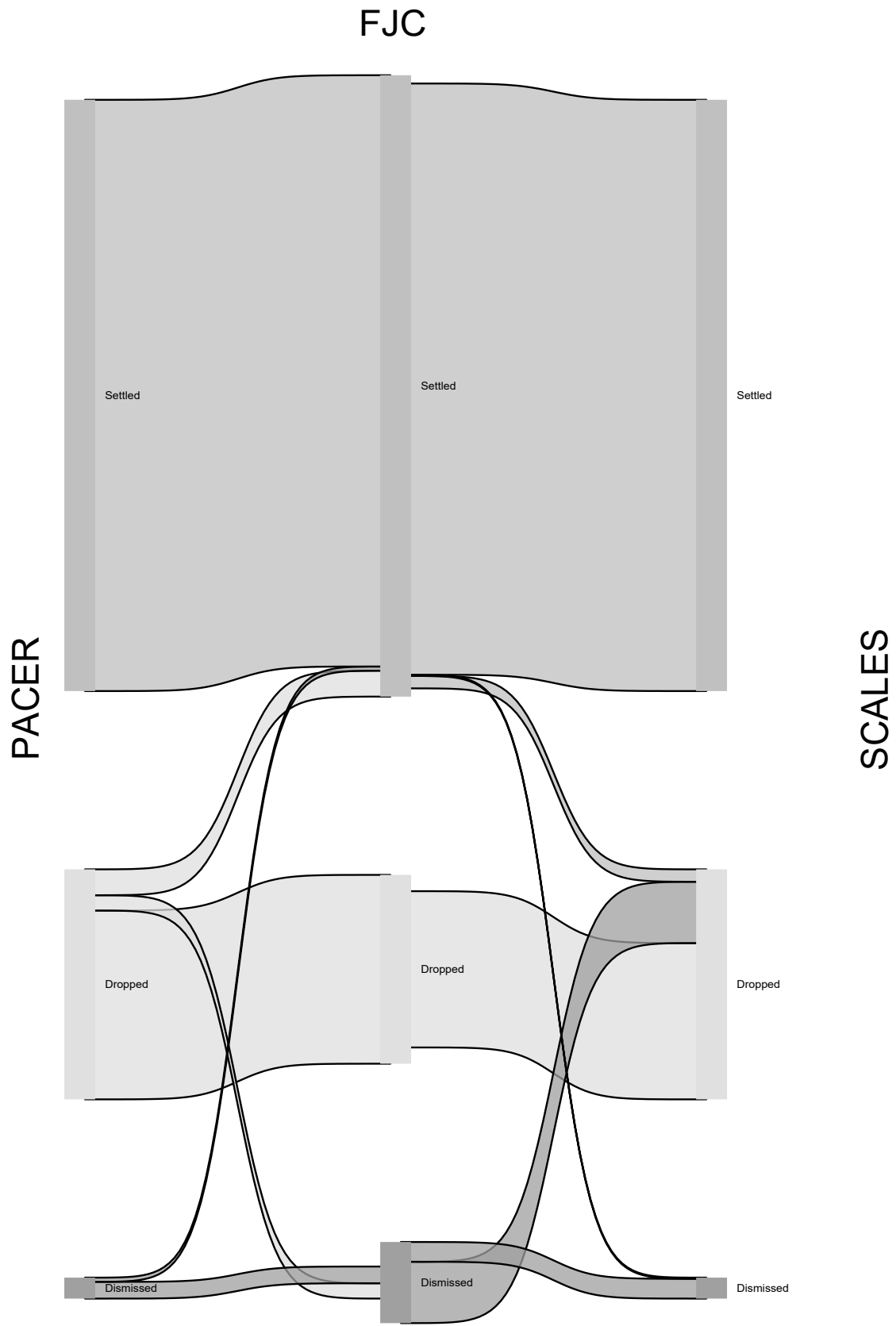

**Figure 3: Impact of Lone Pine Orders and Bellwether Processes (JPML data)**

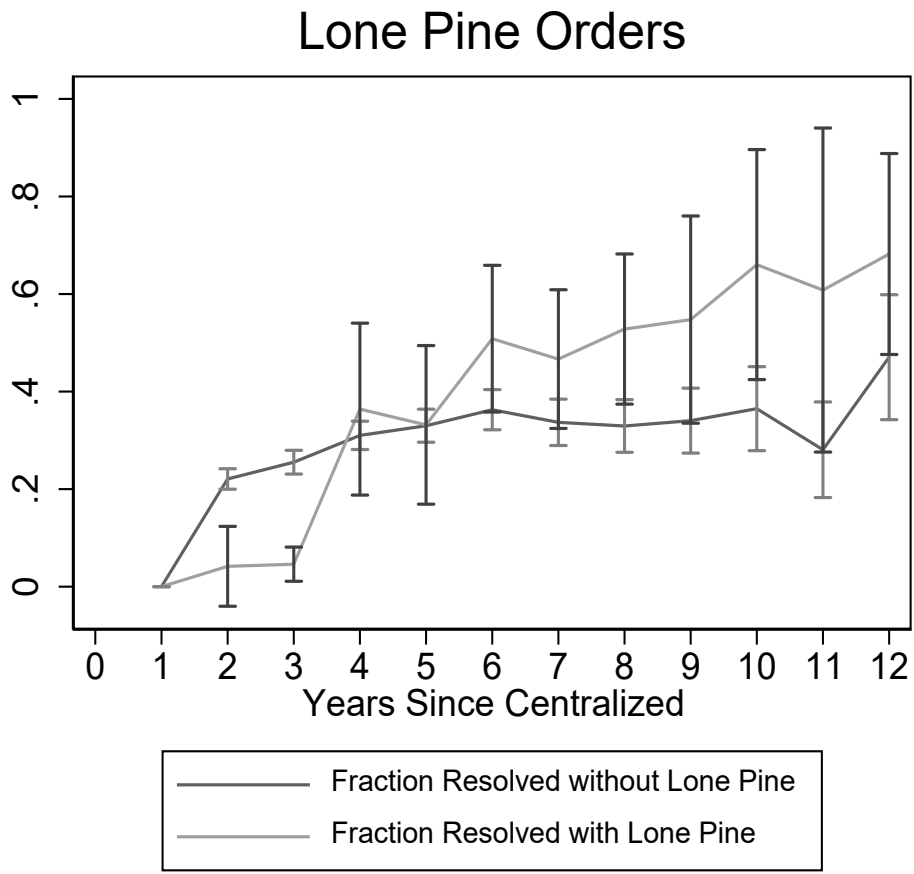

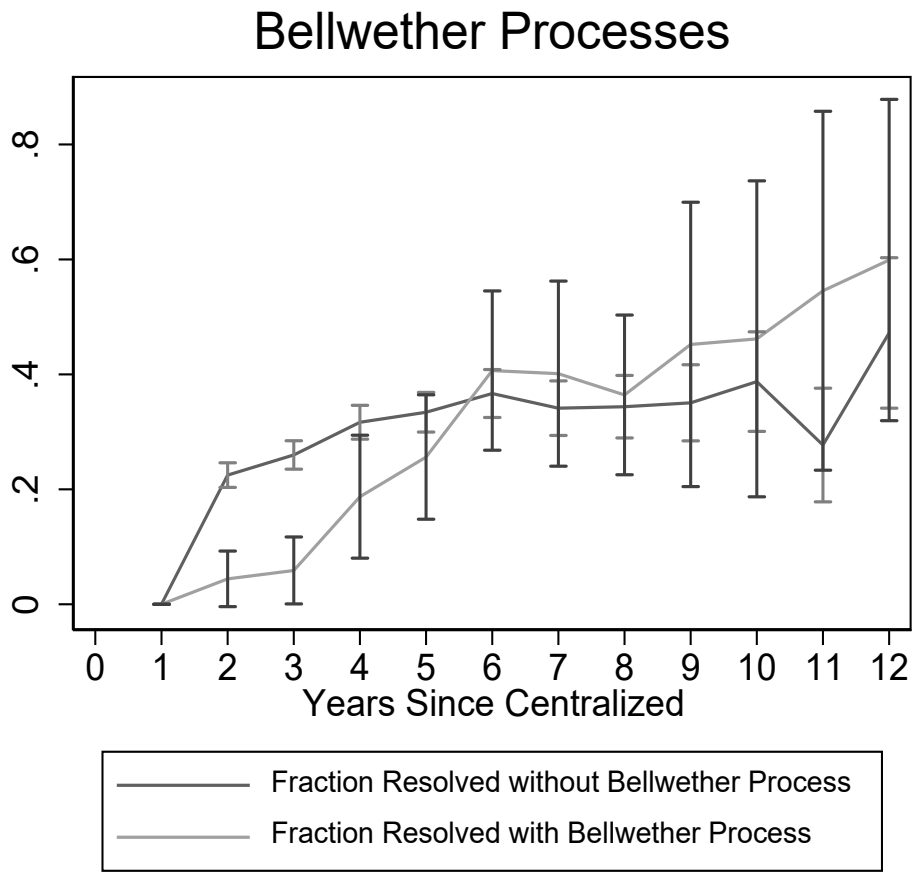

**Figure 4: Monthly Resolution Breakdown Before and After Lone Pine Order in Vioxx MDL**

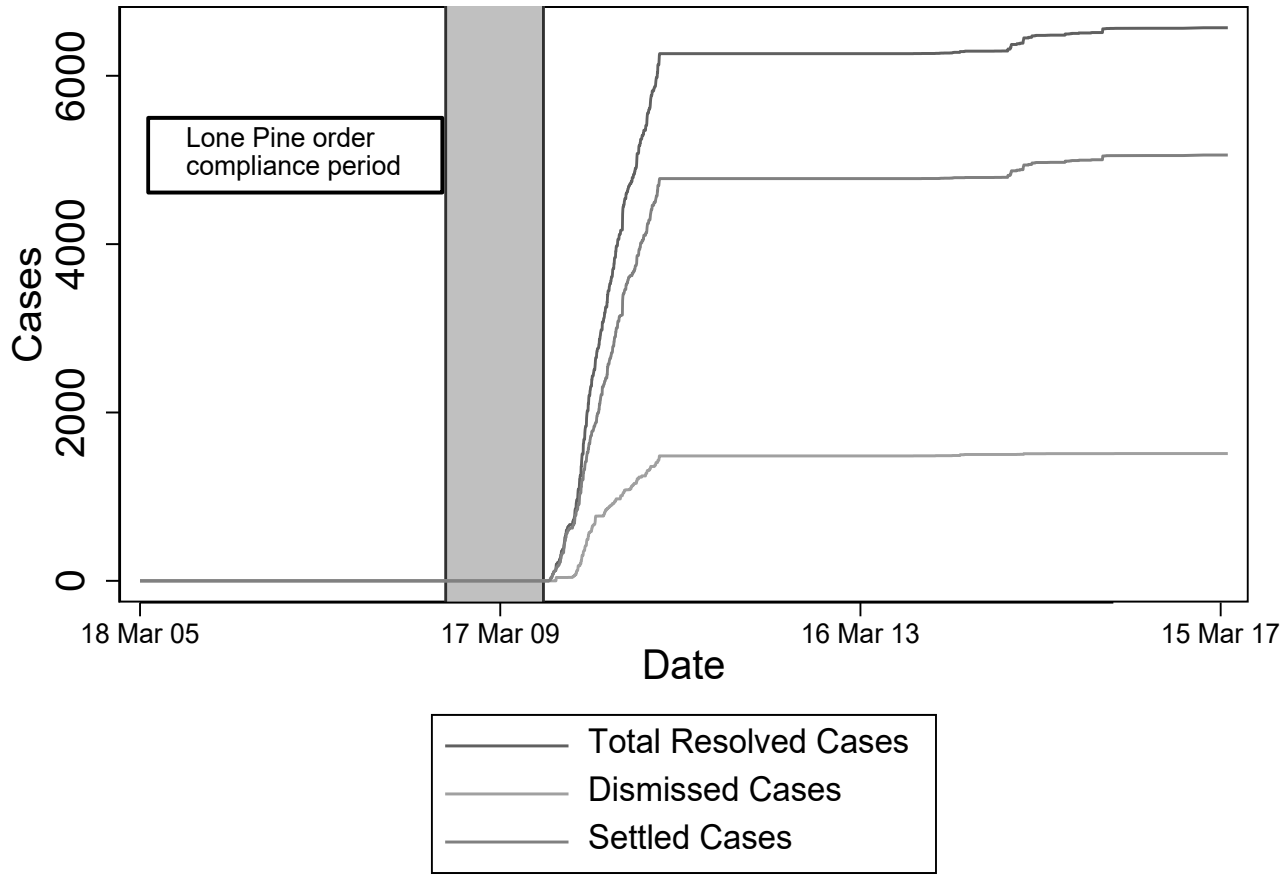

Source: FJC Integrated Data Base Civil Documentation

Appendix Table 1: Use of Plaintiff Fact Sheets Across MDLs

| MDL Number | Case Name | Start of Plaintiff Fact Sheet or Profile | MDL Number | Case Name | Plaintiff Fact |
|---|---|---|---|---|---|
| 1014 | Orthopedic Bone Screw | 1995 | 2325 | American Medical Systems Inc Pelvic Repair System | 2012 |
| 1060 | Baxter Healthcare Gammagard PL | 2003 | 2326 | Boston Scientific Corp Pelvic Repair System | 2012 |
| 1203 | Diet Drugs | 1998 | 2327 | Ethicon Inc Pelvic Repair System | 2012 |
| 1348 | Rezulin | 2000 | 2329 | Wright Medical Technology Inc Conserve Hip Implant | 2012 |
| 1355 | Propulsid | 2001 | 2331 | Propecia (Finasteride) | 2012 |
| 1407 | Phenylpropanolamine (PPA) | 2002 | 2342 | Zoloft (Sertraline Hydrochloriad) | 2012 |
| 1431 | Baycol | 2002 | 2372 | Watson Fentanyl Patch | 2014 |
| 1477 | Serzone | 2003 | 2385 | Pradaxa (Dabigatran Etexilate) | 2012 |
| 1507 | Prempro | 2005 | 2387 | Coloplast Corp Pelvic Support Systems | 2012 |
| 1553 | Silica | 2004 | 2391 | Biomet M2a Hip Implant | 2013 |
| 1626 | Accutane | 2005 | 2404 | Nexium (Esomeprazole) | 2013 |
| 1629 | Neurontin Marketing Sales Practices | 2006 | 2409 | Nexium (Esomeprazole) | 2013 |
| 1657 | Vioxx | 2005 | 2419 | New England Compounding Pharmacy Inc | 2014 |
| 1699 | Bextra and Celebrex Marketing Sales Practices | 2006 | 2428 | Fresenius GranuFlo/NaturaLyte Dialysate | 2013 |
| 1708 | Guidant Corp Implantable Defibrillators | 2006 | 2434 | Mirena IUD | 2013 |
| 1726 | Medtronic Inc Implantable Defibrillators | 2007 | 2436 | Tylenol (Acetaminophen) Marketing Sales Practices and | 2014 |
| 1742 | Ortho Evra | 2009 | 2440 | Cook Medical Inc Pelvic Repair System | 2013 |
| 1760 | Aredia and Zometa | 2007 | 2441 | Stryker Rejuvenate and ABG II Hip Implant | 2013 |
| 1763 | Human Tissue | 2006 | 2452 | Incretin Mimetics | 2013 |
| 1769 | Seroquel | 2007 | 2458 | Effexor (Venlafaxine Hydrochloride) | 2014 |
| 1785 | Bausch and Lomb Inc Contact Lens Solution | 2006 | 2472 | Loestrin 24 Fe | 2014 |
| 1789 | Fosamax | 2007 | 2502 | Lipitor (Atorvastatin Calcium) Marketing Sales Practices and (No II) | 2014 |
| 1836 | Mirapex | 2007 | 2543 | General Motors LLC Ignition Switch | 2014 |
| 1842 | Kugel Mesh Hernia Patch | 2008 | 2545 | Testosterone Replacement Therapy | 2014 |
| 1845 | ConAgra Peanut Butter | 2008 | 2551 | National Hockey League Players' Concussion Injury | 2015 |
| 1871 | Avandia Marketing Sales Practices and | 2008 | 2570 | Cook Medical Inc IVC Filters Marketing Sales Practices and | 2015 |
| 1873 | FEMA Trailer Formaldehyde | 2008 | 2591 | Syngenta AG MIR162 Corn | 2015 |
| 1909 | Gadolinium Contrast Dyes | 2008 | 2592 | Xarelto (Rivaroxaban) | 2015 |
| 1928 | Trasylol | 2008 | 2606 | Benicar (Olmesartan) | 2015 |
| 1943 | Levaquin | 2010 | 2641 | Bard IVC Filters | 2015 |
| 1953 | Heparin | 2009 | 2642 | Fluoroquinolone | 2016 |
| 1964 | NuvaRing | 2009 | 2652 | Ethicon Inc Power Morcellator | 2016 |
| 1968 | Digitek | 2009 | 2654 | Amtrak Train Derailment in Philadelphia PA on 5/12/15 | 2017 |
| 2004 | Mentor Corp ObTape Transobturator Sling | 2012 | 2657 | Zofran (Ondansetron) | 2016 |
| 2016 | Yamaha Motor Corp RhinoV | 2010 | 2666 | Bair Hugger Forced Air Warming Devices | 2016 |
| 2047 | ChineseManufactured Drywall | 2009 | 2672 | Volkswagen Clean Diesel Marketing Sales Practices and | 2017 |
| 2051 | Denture Cream | 2009 | 2687 | Liquid Aluminum Sulfate | 2016 |
| 2066 | Oral Sodium PhosphateBased | 2009 | 2691 | Viagra (Sildenafil Citrate) and Cialis (Tadalafil) | 2017 |
| 2084 | AndroGel (No II) | 2014 | 2734 | Abilify (Aripiprazole) | 2017 |
| 2087 | Hydroxycut Marketing and Sales Practices | 2010 | 2738 | Johnson and Johnson Talcum Powder Products Marketing Sales Practices and | 2017 |
| 2092 | Chantix (Varenicline) | 2011 | 2740 | Taxotere (Docetaxel) | 2017 |
| 2100 | Yasmin and Yaz (Drospirenone) Marketing Sales Practices | 2011 | 2750 | Invokana (Canagliflozin) | 2017 |
| 2151 | Toyota Motor Corp Unintended Acceleration Marketing Sales Practices and | 2010 | 2753 | Atrium Medical Corp CQur Mesh | 2017 |
| 2158 | Zimmer Durom Hip Cup | 2013 | 2754 | Eliquis (Apixaban) | 2017 |
| 2179 | Oil Spill by the Oil Rig Deepwater Horizon in the Gulf of Mexico 4/20/2010 | 2010 | 2767 | Mirena IUS LevonorgestrelRelated (No II) | 2017 |
| 2187 | CR Bard Inc Pelvic Repair Systems | 2011 | 2768 | Stryker LFIT V40 Femoral Head | 2017 |
| 2197 | DePuy Orthopaedics Inc ASR Hip Implant | 2011 | 2776 | Farziga (Dapagliflozin) | 2017 |
| 2243 | Fosamax (Alendronate Sodium (No II) | 2011 | 2782 | Ethicon Physiomesh Flexible Composite Hernia Mesh | 2017 |
| 2272 | Zimmer NexGen Knee Implant | 2012 | 2785 | EpiPen (Epinephrine Injection USales Practices) Marketing Sales Practices and | 2017 |
| 2299 | Actos (Pioglitazone) | 2012 | 2789 | ProtonPump Inhibitor (No II) | 2017 |
| 2308 | Skechers Toning Shoe | 2012 | | | |

**Appendix Table 2: Use of Defendant Fact Sheets Across MDLs**

| MDL Number | Case Name | Start of Defendant Fact Sheet |
|---|---|---|
| 1553 | Silica | 2004 |
| 1657 | Vioxx | 2007 |
| 1699 | Bextra and Celebrex Marketing Sales Practices | 2006 |
| 1726 | Medtronic Inc Implantable Defibrillators | 2007 |
| 1842 | Kugel Mesh Hernia Patch | 2008 |
| 1909 | Gadolinium Contrast Dyes | 2008 |
| 1943 | Levaquin | 2010 |
| 1953 | Heparin | 2009 |
| 1968 | Digitek | 2009 |
| 2004 | Mentor Corp ObTape Transobturator Sling | 2016 |
| 2047 | ChineseManufactured Drywall | 2009 |
| 2187 | CR Bard Inc Pelvic Repair Systems | 2011 |
| 2197 | DePuy Orthopaedics Inc ASR Hip Implant | 2011 |
| 2299 | Actos (Pioglitazone) | 2012 |
| 2325 | American Medical Systems Inc Pelvic Repair System | 2012 |
| 2326 | Boston Scientific Corp Pelvic Repair System | 2012 |
| 2327 | Ethicon Inc Pelvic Repair System | 2012 |
| 2329 | Wright Medical Technology Inc Conserve Hip Implant | 2012 |
| 2342 | Zoloft (Sertraline Hydrochloriad) | 2012 |
| 2385 | Pradaxa (Dabigatran Etexilate) | 2012 |
| 2387 | Coloplast Corp Pelvic Support Systems | 2012 |
| 2391 | Biomet M2a Hip Implant | 2014 |
| 2428 | Fresenius GranuFlo/NaturaLyte Dialysate | 2013 |
| 2440 | Cook Medical Inc Pelvic Repair System | 2013 |
| 2441 | Stryker Rejuvenate and ABG II Hip Implant | 2013 |
| 2452 | Incretin Mimetics | 2013 |
| 2472 | Loestrin 24 Fe | 2014 |
| 2502 | Lipitor (Atorvastatin Calcium) Marketing Sales Practices and (No II) | 2014 |
| 2551 | National Hockey League Players' Concussion Injury | 2015 |
| 2570 | Cook Medical Inc IVC Filters Marketing Sales Practices and | 2015 |
| 2592 | Xarelto (Rivaroxaban) | 2015 |
| 2606 | Benicar (Olmesartan) | 2015 |
| 2641 | Bard IVC Filters | 2015 |
| 2642 | Fluoroquinolone | 2016 |
| 2652 | Ethicon Inc Power Morcellator | 2016 |
| 2657 | Zofran (Ondansetron) | 2016 |
| 2691 | Viagra (Sildenafil Citrate) and Cialis (Tadalafil) | 2017 |
| 2734 | Abilify (Aripiprazole) | 2017 |
| 2740 | Taxotere (Docetaxel) | 2017 |
| 2750 | Invokana (Canagliflozin) | 2017 |
| 2753 | Atrium Medical Corp CQur Mesh | 2017 |
| 2754 | Eliquis (Apixaban) | 2017 |
| 2768 | Stryker LFIT V40 Femoral Head | 2017 |
| 2782 | Ethicon Physiomesh Flexible Composite Hernia Mesh | 2017 |
| 2789 | ProtonPump Inhibitor (No II) | 2017 |